\documentclass[jour nal]{IEEEtran}
\usepackage{booktabs}
\usepackage{amsmath,amssymb}
\usepackage{cases}
\usepackage{tabularx}
\usepackage{multicol}
\usepackage{graphicx}
\usepackage{float}
\usepackage{amsmath}
\usepackage{amssymb}
\usepackage{array}
\usepackage{color}
\usepackage{xcolor}
\usepackage{multirow}
\usepackage{color}
\usepackage{algorithmic}
\usepackage{bbm}
\usepackage{bm}
\usepackage{tikz}
\newcommand{\specialcell}[2][c]{%
  \begin{tabular}[#1]{@{}c@{}}#2\end{tabular}}
\graphicspath{{figures/}}
\usepackage[numbers,sort&compress]{natbib}

\usepackage[bookmarks,colorlinks,linkcolor=red, anchorcolor=blue, citecolor=green]{hyperref}
\usepackage{url}
\newcommand{\etal}{\textit{et al}.}
\newcommand{\ie}{\textit{i}.\textit{e}.}

\newcommand{\GRP}{\mathrm{GP}}

\graphicspath{{figs/}}

\begin{document}

\title{Learning Context-Based Non-local Entropy Modeling for Image Compression}

\author{Mu~Li,
        Kai~Zhang,
        Wangmeng~Zuo,~\IEEEmembership{Senior Member,~IEEE},
        Radu~Timofte,~\IEEEmembership{Member,~IEEE},
        and~David~Zhang,~\IEEEmembership{Fellow,~IEEE}
\IEEEcompsocitemizethanks{
\IEEEcompsocthanksitem Mu Li is with the School of Science and Engineering, The Chinese University of Hong Kong, Shenzhen, Guangdong, 518172, P.R. China (e-mail: limuhit@gmail.com).
\IEEEcompsocthanksitem Kai Zhang and Radu Timofte are with the Computer Vision Laboratory, ETH Zurich, Zurich, Switzerland.
\IEEEcompsocthanksitem Wangmeng Zuo is with the School of Computer Science and Technology, Harbin Institute of Technology, Harbin, 150001, China (e-mail: cswmzuo@gmail.com).
\IEEEcompsocthanksitem David Zhang is with the School of Science and Engineering, The Chinese University of Hong Kong (Shenzhen),  Shenzhen Research Institute of Big Data, and Shenzhen Institute of Artificial Intelligence and Robotics for Society, Shenzhen, China, e-mail: (davidzhang@cuhk.edu.cn).}
}

\markboth{Submitted to IEEE Transactions on Neural Networks and Learning Systems}%
{Shell \MakeLowercase{\textit{et al.}}: Bare Demo of IEEEtran.cls for IEEE Journals}

\maketitle

\begin{abstract}
The entropy of the codes usually serves as the rate loss in the recent learned lossy image compression methods.
Precise estimation of the probabilistic distribution of the codes plays a vital role in the performance.
However, existing deep learning based entropy modeling methods generally assume the latent codes are statistically independent or depend on some side information or local context, which fails to take the global similarity within the context into account and thus hinder the accurate entropy estimation.
To address this issue, we propose a non-local operation for context modeling by employing the global similarity within the context.
Specifically, we first introduce the proxy similarity functions and spatial masks to handle the missing reference problem in context modeling.
%
Then, we combine the local and the global context via a non-local attention block and employ it in masked convolutional networks for entropy modeling.
The entropy model is further adopted as the rate loss in a joint rate-distortion optimization to guide the training of the analysis transform and the synthesis transform network in transforming coding framework.
Considering that the width of the transforms is essential in training low distortion models, we finally produce an U-Net block in the transforms to increase the width with manageable memory consumption and time complexity.
Experiments on Kodak and Tecnick datasets demonstrate the superiority of the proposed context-based non-local attention block in entropy modeling and the U-Net block in low distortion compression against the existing image compression standards and recent deep image compression models.

\end{abstract}


%
\IEEEpeerreviewmaketitle

\section{Introduction}
Image compression is a crucial problem in computer science that has been studied for decades. Many image compression standards like JPEG~\cite{wallace1992jpeg} have already been used in our daily life. Nevertheless, the last decade has witnessed a population of artificial intelligence and social media, which brought new challenges for sharing and storing huge amounts of high-definition media. Better image compression methods are still demanding.

Deep neural networks (DNNs) have proven to be effective and get extraordinary results in numerous vision tasks such as image restoration~\cite{zhang2017beyond,dong2014learning,dong2016image,majumdar2018blind}, image quality assessment~\cite{ma2018,wang2015blind}, and image classification~\cite{tang2018deep,luo2017convolutional}, which throws light on learning better lossy image compression methods.
Deep neural networks are natural transforms. As a result, nearly all of the recent attempts on image compression with DNNs follow the transforming coding framework and are optimized for a rate-distortion trade-off.  A transforming coding
framework usually consists of three components, i.e., transform, quantization and entropy coding. To be specific, transform maps the input images to code representations; quantization transforms the representation into discrete counterparts;  entropy coding compresses the quantized codes into the bitstream in a lossless manner.

Previous transforms~\cite{ahmed_1974_dct} are linear, invertible and fixed for all bit rates. Distortion only arises from the quantization.
As for the recent deep lossy image compression methods, transforms are modeled by learning non-linear and powerful DNNs. 
%
With the quantization operation and the structure of the network, the learned transforms are generally non-invertible, which
tends to encourage discarding the perceptual negligible information
for better visual quality at low bit rates.  However,
the performance is usually limited at high bit rates where the distortion is required to be as small as possible. Empirically, the width of the transforms, i.e., the number of feature maps at each layer, is effective in reducing the information losing brought by the transforms.
Many deep image compression methods~\cite{rippel2017real,balle2018variational,minnen2018joint} suggest adopting wider transforms at high bit rates.
For a deep network, the growing of width will inevitably increase the computational complexity and GPU memory consumption.
We suggest adopting an U-Net like block in the transforms which could help reduce the time complexity and memory usage in the transforms.  With the paired down-sampling, up-sampling operations and skip connection, the U-Net like architecture
can not only speed up the transforms but also combine the information in different scales and facilitate the information propagation.
%

The learned image compression methods are usually modeled as a rate-distortion optimization problem where entropy modeling of the codes plays an important role. According to Shannon's source coding theorem~\cite{Shannon1948}, given a sequence of codes $\bm y =\{y_0,\ldots,y_M\}$, the optimal code length of $\bm y$ should be $\lceil-\log_{2}P(\bm y)\rceil$ with the codebook constructed by binary symbols. Thus, building accurate discrete probability distribution functions for the codes is essential in determining the compression rate. Most entropy coding schemes directly assume the codes in $\bm y$ are independent and follow the same marginal distribution, resulting in a code length of $\lceil-\sum_{i=0}^{M}\log_{2}P(y_i)\rceil$. Generally, a better approximation could be given with the chain rule when considering the context.
\begin{align}
P(\bm y) \approx \prod_{i=0}^{M}P(y_i|y_0,\ldots,y_{i-1}),
\end{align}
where the context of $y_i$ is defined as all the codes coded before it in $\bm y$.
One typical application of the context in entropy modeling is the context-based adaptive binary arithmetic coding (CABAC)~\cite{marpe2003context} used in H.264/AVC, which gets a clear performance improvement over previous image compression standards by considering only two nearest codes in the context. However, the traditional context-based entropy modeling based on histograms fails to cover more codes.
To employ more codes in the context and the shared computation of fully convolutional network, a set of masked convolutional networks including PixelCNN~\cite{minnen2018joint}, MCN~\cite{mentzer2018conditional1}, and CCN~\cite{li2019efficient} are proposed for entropy modeling in recent lossy image compression methods.
Nevertheless, all the entropy models based on CNNs have a limited receptive filed and only consider the local context within the receptive field in entropy modeling. Global information and content similarity in the context are ignored.

To explore global information, we introduce a context-based non-local operation for entropy modeling.
In image denoising, non-local means~\cite{buades2005non} predicts the target pixel with a weighted sum on the whole image, where the weights are produced by the content similarity between the target pixel and the others.
Inspired by the non-local means, we regress a target code, $y_r(p,q)$, with a weighted sum over all the codes in the $r$-th channel within the context.
Here, $y_r(p,q)$ is a code from the 3D code block $\bm y$ at the position $(p,q)$ in the $r$-th channel.
However, due to the definition of context, the target code itself is unknown in the calculation of the weights.
Without the reference, it is unable to evaluate the similarity and produce the weights.
We call this challenge missing reference problem.
Considering that $y_r(p,q)$ and the codes from the vector $\bm \xi_r(p,q)=\{y_0(p,q),\ldots,y_{r-1}(p,q)\}$ are produced by the same image patch, there should be a close relationship between them. We alternatively evaluate the similarity between the target code $y_r(p,q)$ and $y_r(u,v)$ as a weighted L2 distance between $\bm \xi_r(p,q)$ and $\bm \xi_r(u,v)$.
An attention block is introduced to combine the local representations produced by CCNs and the global representations from the non-local operation for context-based entropy modeling.

The context-based non-local attention block is further adopted in CCNs to estimate the entropy of the lossy image compression method. In detail, the probability of each code is modeled on a mixture of Gaussian (MoG) distribution whose parameters are predicted by the non-local CCNs.
We jointly optimize transforms, \ie, analysis transform and synthesis transform built on U-Net blocks and the context-based non-local entropy model with respect to the trade-off of rate-distortion performance in an end-to-end manner.
Experiments on the Kodak and Tecnick datasets show that the proposed method can outperform state-of-art lossy image compression standards.

\section{Related Work}

\subsection{DNN-Based Lossy Image Compression Methods}
Recent DNN-based lossy image compression methods are optimized for the rate-distortion trade-off, where the rate loss is usually modeled as the entropy of the codes. Previous methods~\cite{balle2016end,theis2017lossy,agustsson2017soft,rippel2017real} only adopt DNNs for transforms and suppose that all the codes are i.i.d. and follow the same probability distribution function (PDF) for easy entropy models. With a lack of accurate estimation of the entropy, the performance of such methods is limited. Arguably speaking, the code representations generated from the image via a highly nonlinear analysis transform still exhibit strong statistical redundancies~\cite{balle2018variational}. Therefore, several attempts have been made by estimating the conditional PDFs of the codes based on extra information, \ie, context~\cite{toderici2016full,li2017learning,minnen2018joint,li2019efficient,mentzer2018conditional1} and hyperprior~\cite{balle2018variational,minnen2018joint}.
Specially, for context-based entropy modeling, DNN-based entropy models such as RNN~\cite{mikolov2010recurrent}, LSTM~\cite{sundermeyer2012lstm} are shown to be powerful in modeling the context of sequential data in natural language processing.
PixelRNN~\cite{oord2016pixel} and PixelCNN~\cite{oord2016conditional} are further proposed for modeling long-range dependency among pixels and employ much larger context for processing image data.
Taking both of the context and effeciency into account, a set of masked convolutional networks, such as PixelCNN~\cite{minnen2018joint}, MCN~\cite{mentzer2018conditional1}, and CCN~\cite{li2019efficient}, are proposed for modeling the entropy of the codes in lossy image compression.
In the following, we will give an overview of some representative DNN-based lossy image compression methods.

\begin{figure*}[!htbp]
	\centering
	\begin{minipage}{1.0\linewidth}
	\begin{tikzpicture}
    	\node[inner sep=0pt] (nd1) at (0,0)
    	{\includegraphics[width=1.0\linewidth]{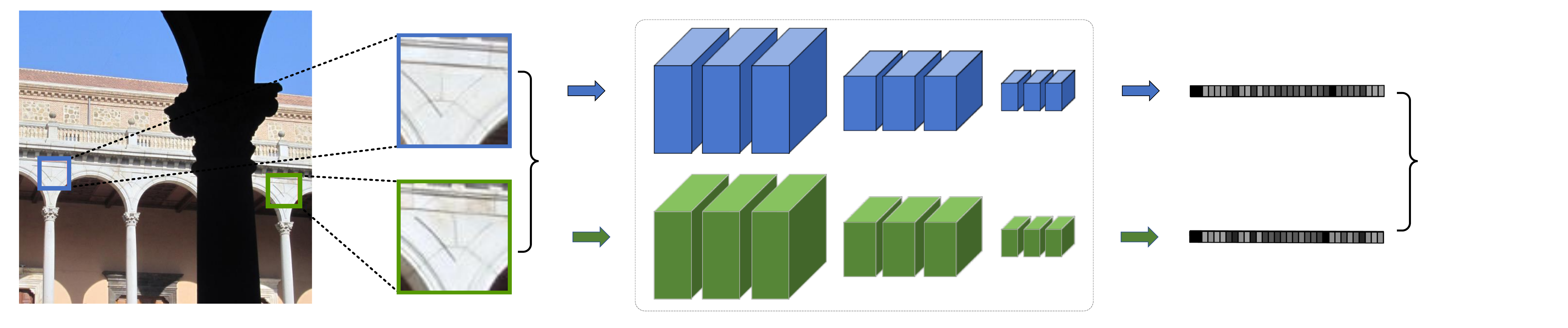}};
		\node[text width=1.5cm,align=center] (cp1) at (-2.3,-0.1){\baselineskip=3pt \scriptsize{Similar patches} \par};
		\node[text width=1.5cm,align=center] (cp1) at (8,0){\baselineskip=3pt \scriptsize{Similar code vectors} \par};
		\node[text width=3.5cm,align=center] (cp1) at (-7.2,-1.9){\baselineskip=3pt \scriptsize{Source image} \par};
		\node[text width=3.5cm,align=center] (cp1) at (1.2,-1.9){\baselineskip=3pt \scriptsize{Analysis Transform} \par};
		\node[text width=3.5cm,align=center] (cp1) at (5.9,-1.9){\baselineskip=3pt \scriptsize{Code vector} \par};
	\end{tikzpicture}
	\end{minipage}
	
	\hfill\vline\hfill

	\begin{minipage}{1.0\linewidth}
		\centering{\scriptsize{(a)}}
	\end{minipage}

	\hfill\vline\hfill
	
	\begin{minipage}{1.0\linewidth}
	\begin{minipage}{0.3\linewidth}
	\begin{minipage}{1.0\linewidth}
	\begin{tikzpicture}
    	\node[inner sep=0pt] (nd1) at (0,0)
    	{\includegraphics[width=1.0\linewidth]{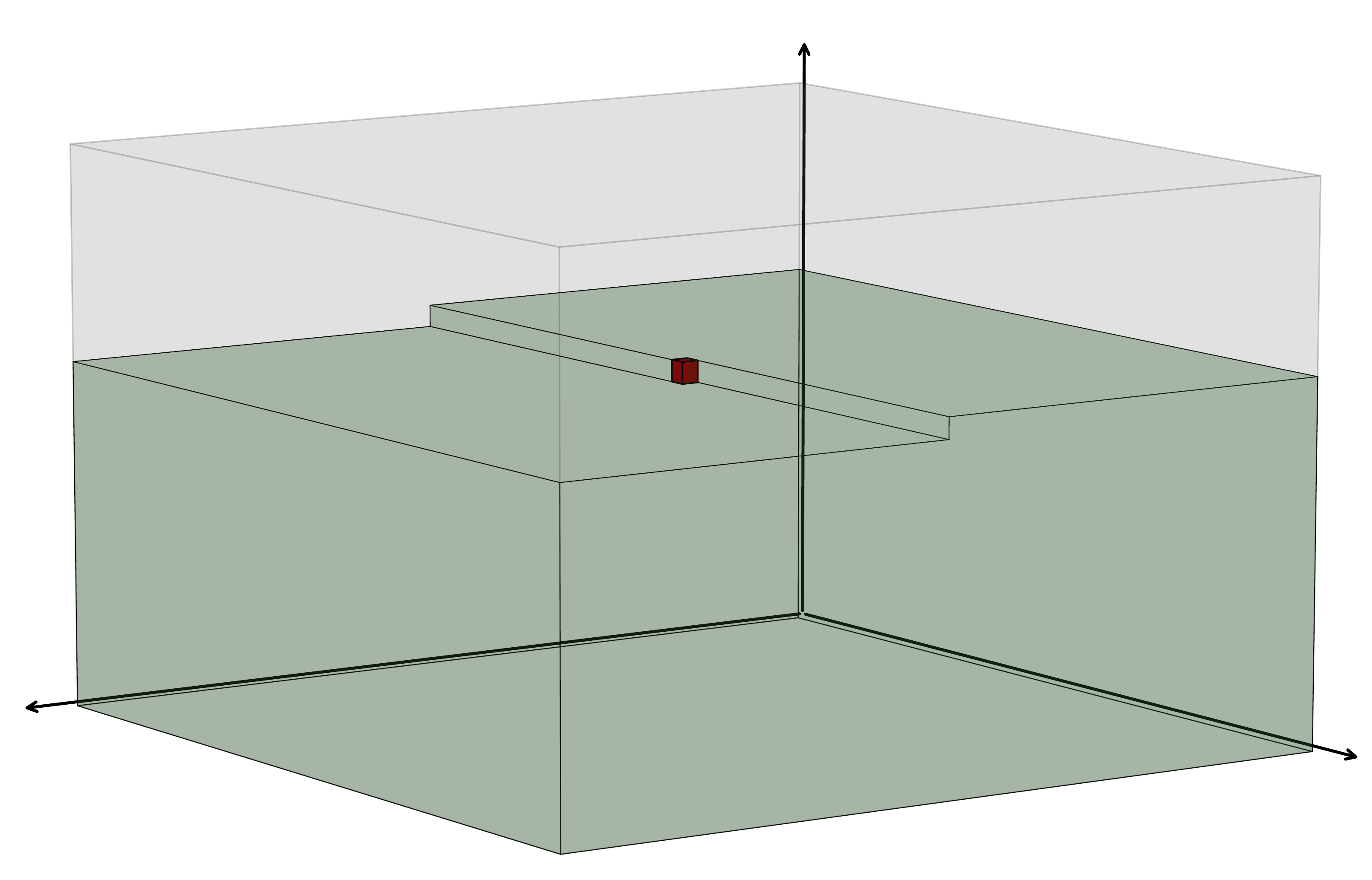}};
		\draw [line width=0.25mm,densely dotted] (0,0.3) -- (2,1.5);
		\node[text width=2cm,align=center] (cp1) at (2,1.8){\baselineskip=3pt \scriptsize{Target code} \par};
		\draw [line width=0.25mm,densely dotted] (0,-0.6) -- (-1.5,1.6);
		\node[text width=2cm,align=center] (cp2) at (-1.5,1.8){\baselineskip=3pt \scriptsize{Full context} \par};
		\node[text width=0.25cm,align=center] (cp3) at (-2.6,-0.9){\baselineskip=3pt {$p$} \par};
		\node[text width=0.25cm,align=center] (cp4) at (2.7,-1.4){\baselineskip=3pt {$q$} \par};
		\node[text width=0.25cm,align=center] (cp5) at (0.5,1.75){\baselineskip=3pt {$r$} \par};
	\end{tikzpicture}
	\end{minipage}
	
	\hfill\vline\hfill
	\hfill\vline\hfill
	
	\begin{minipage}{1.0\linewidth}
		\centering{\scriptsize{(b)}}
	\end{minipage}
	

	\end{minipage}
	\begin{minipage}{0.3\linewidth}
	\begin{minipage}{1.0\linewidth}
	\begin{tikzpicture}
    	\node[inner sep=0pt] (nd1) at (0,0)
    	{\includegraphics[width=1.0\linewidth]{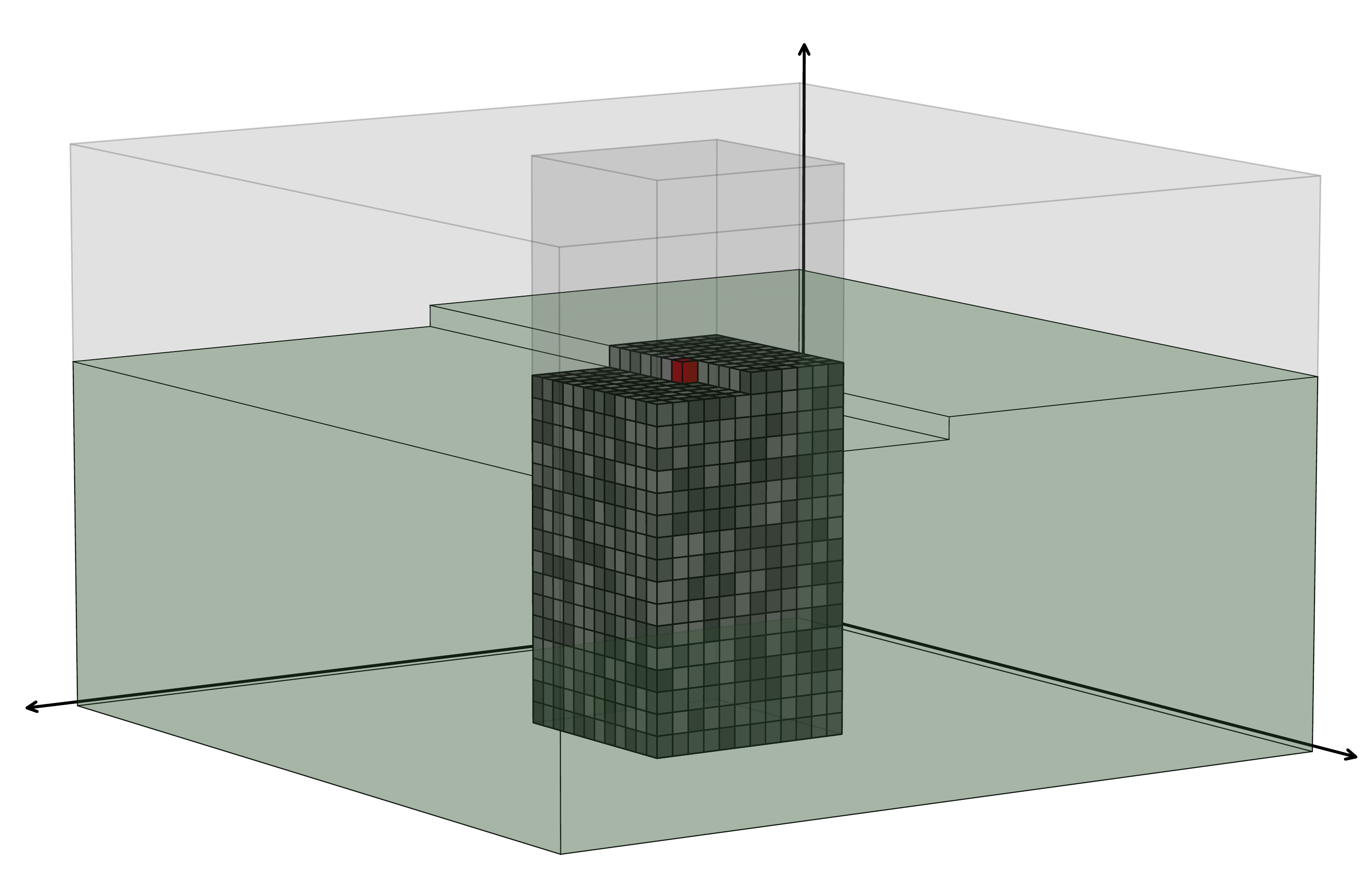}};
		\draw [line width=0.25mm,densely dotted] (0,0.3) -- (2,1.5);
		\node[text width=2cm,align=center] (cp1) at (2,1.8){\baselineskip=3pt \scriptsize{Target code} \par};
		\draw [line width=0.25mm,densely dotted] (0,1) -- (-0.5,1.6);
		\node[text width=2cm,align=center] (cp2) at (-0.5,1.8){\baselineskip=3pt \scriptsize{Local context} \par};
		\node[text width=0.25cm,align=center] (cp3) at (-2.6,-0.9){\baselineskip=3pt {$p$} \par};
		\node[text width=0.25cm,align=center] (cp4) at (2.7,-1.4){\baselineskip=3pt {$q$} \par};
		\node[text width=0.25cm,align=center] (cp5) at (0.5,1.75){\baselineskip=3pt {$r$} \par};
	\end{tikzpicture}
	\end{minipage}
	
	\hfill\vline\hfill
	\hfill\vline\hfill
	
	\begin{minipage}{1.0\linewidth}
		\centering{\scriptsize{(c)}}
	\end{minipage}
	
	
	\end{minipage}
	\begin{minipage}{0.3\linewidth}
	\begin{minipage}{1.0\linewidth}
	\begin{tikzpicture}
    	\node[inner sep=0pt] (nd1) at (0,0)
    	{\includegraphics[width=1.0\linewidth]{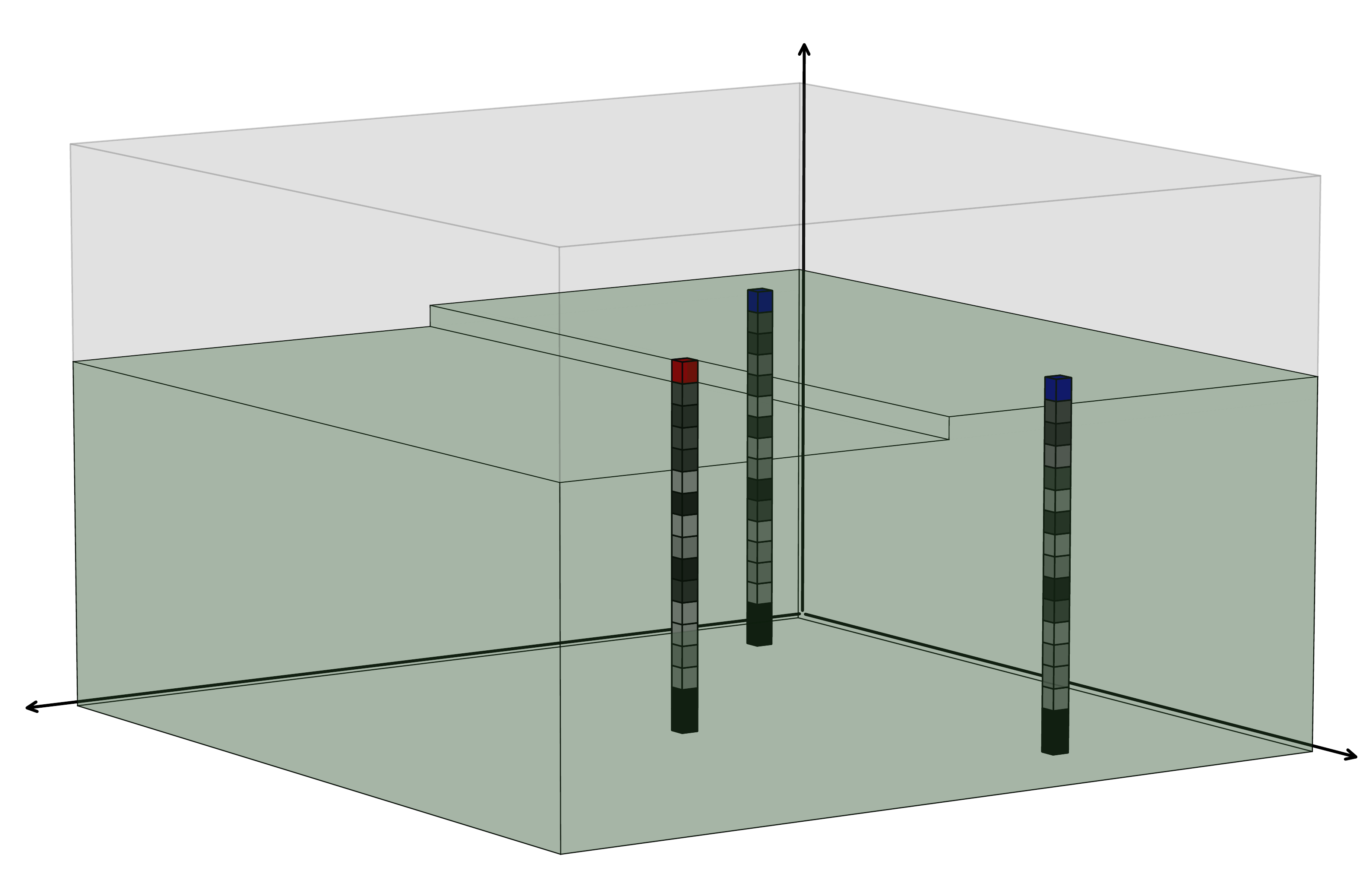}};
		\draw [line width=0.25mm,densely dotted] (0,0.3) -- (2,1.5);
		\node[text width=2cm,align=center] (cp1) at (2,1.8){\baselineskip=3pt \scriptsize{Target code} \par};
		\draw [line width=0.25mm,densely dotted] (0,-0.6) -- (3,-0.6);
		\draw [line width=0.25mm,densely dotted] (0.3,-0.3) -- (3,-0.6);
		\draw [line width=0.25mm,densely dotted] (1.45,-0.7) -- (3,-0.6);
		\node[text width=2cm,align=center] (cp2) at (3.5,-0.6){\baselineskip=3pt \scriptsize{Similar code vectors} \par};
		\node[text width=0.25cm,align=center] (cp3) at (-2.6,-0.9){\baselineskip=3pt {$p$} \par};
		\node[text width=0.25cm,align=center] (cp4) at (2.7,-1.4){\baselineskip=3pt {$q$} \par};
		\node[text width=0.25cm,align=center] (cp5) at (0.5,1.75){\baselineskip=3pt {$r$} \par};
	\end{tikzpicture}
	\end{minipage}
	
	\hfill\vline\hfill
	\hfill\vline\hfill
	
	\begin{minipage}{1.0\linewidth}
		\centering{\scriptsize{(d)}}
	\end{minipage}
	
	
	\end{minipage}
	
	\end{minipage}
	
	\caption{Illustration of context-based non-local entropy modeling. (a) indicates the non-local similarity among the codes generated by the analysis transform. (b) shows the global context of a target code in 3D code block $\bm y$ with a raster scanning order. (c) gives the local context used in CNN-based entropy modeling. (d) illustrates the non-local similarity in the context.}
	\label{fig:intro}
\end{figure*}
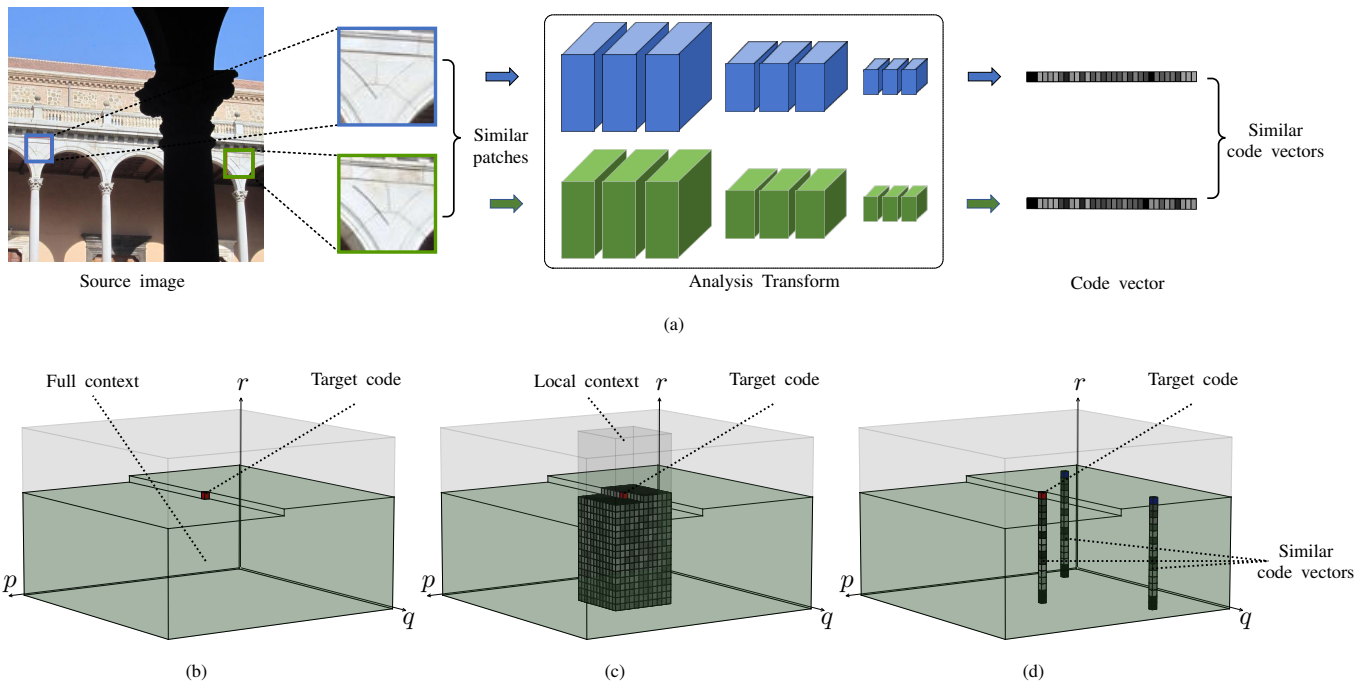

Toderici \etal~\cite{toderici2015variable} proposed an RNN to compress $32\times 32$ images in a progressive manner. They later extended the job to full-size images and introduced a BinaryRNN for context-based entropy modeling~\cite{toderici2016full}. Johnston \etal~\cite{johnston2017improved} further introduced content-adaptive bit allocation, warm-start training tricks, and perceptual losses to boost the compression performance in terms of MS-SSIM.

Ball{\'e} \etal~\cite{balle2016end} proposed to learn shallow transforms with a GDN activation function in an end-to-end manner and model the entropy with a shared linear piece-wise PDF.
In their following work~\cite{balle2018variational}, each code is supposed to follow a zero-mean Gaussian distribution with the deviation estimated from a side information network depending on the hierarchical hyperprior.
Minnen \etal~\cite{minnen2018joint} combined the hierarchical hyperprior with context-based auto-regressive prior to boosting the compression performance.

Theis \etal~\cite{theis2017lossy} proposed a strait-forward relaxation of the quantization operations and exploited the Gaussian scale mixture for entropy modeling. Rippel \etal~\cite{rippel2017real} proposed a pyramid-based network structure for analysis and synthesis transforms and an adaptive code length regularization for real-time image compression. And a generative adversarial loss~\cite{goodfellow2014generative} was introduced to generate visually better decoded images at low bit rates. Later, Agustsson \etal~\cite{agustsson2017soft} proposed a soft-to-hard quantization scheme with a parametric softmax function.

Li \etal~\cite{li2017learning} learned an importance map as the side information for content variant rate controlling and exploited a simple CNN for context-based entropy modeling.
Mentzer \etal~\cite{mentzer2018conditional1} further proposed a mask convolutional network for context-based entropy modeling and alternatively optimized the entropy model, transforms, and the importance map.
Limited by the raster coding order,  the context-based models should be decoded in serial coding order and thus are computational inefficiency.
The heavy computational burden makes them not practical in many real applications.
To speed up the entropy modeling and employ large context, context-based convolutional networks (CCNs)~\cite{li2019efficient} introduce a specially defined coding order and group context for entropy modeling in both lossy and lossless image compression. With a given code dividing scheme, the CCNs can perform parallel computation within a group in decoding and largely speed up the decoding speed.

%

Previous deep context-based entropy models only focus on local context in the receptive field of CNNs and usually ignore the non-local similarity in the global context. To tackle this issue, we take the non-local similarity within the global context into account and combine it with the local representation for entropy modeling.

\subsection{Non-local Methods for Image Processing}
Non-local methods were first proposed for employing the global similarity among the pixels in image denoising. Later, DNN-based image processing methods embed it into DNNs to employ the global information and boost the performance in different tasks.
Buades \etal~\cite{buades2005non} explored the self-similarity among pixels and proposed the non-local means based on a content weighted non-local averaging of all pixels in the image for image denoising. Wang \etal~\cite{wang2018non} formulated the non-local operation as a uniform block and adopt it in DNNs to combine local and non-local information for object detection. Liu \etal~\cite{liu2018non} proposed a non-local recurrent network which incorporates non-local operations into a recurrent network for image restoration. Zhang \etal~\cite{zhang2019residual} exploited the non-local operation to building attention masks to capture long-range dependency between pixels and pay more attention to the challenging parts for high-quality image restoration.
In this paper, we tackle the missing reference problem in context modeling with the proposed proxy similarity metric and introduce the non-local operation for the context-based entropy modeling. And a non-local attention block is introduced to combine local and global context for more accurate entropy estimation.

\section{Context-based Entropy Modeling}

Modeling the entropy of the code from its context is an auto-regression problem where the probability distribution functions (PDFs) of a code is regressed from the context of it. Let $\bm y\in\mathbb{R}^{M\times H\times W}$ denote a 3D code block generated by the analysis transform, where $M$, $H$, and $W$ separately are the channel, height, and width of the code block. $y_r(p,q)$ is a code at the position $(p,q)$ in $r$-th channel. With a given coding order, the context of a target $y_r(p,q)$ among a code block $\bm y$ is defined as all the codes scanned before it, \ie, $\mathrm{CTX}(y_r(p,q),\bm y)$.  Fig.~\ref{fig:intro} (b) gives an example of the context of the target code in red with a raster scanning order.

Thanks to the CNN's efficiency in processing 2D and 3D media data, previous context-based entropy modeling usually adopts masked CNNs to predict the PDFs of the codes. As shown in Fig.~\ref{fig:intro} (c), limited by the structure, CNNs only focus on the local context surrounding the target code in the receptive field. The content similarity and global context are generally ignored. To exploit the global information and the content similarity between the target code and its context, we introduce the non-local operation for context-based entropy modeling.

Non-local operation is usually defined as a weighted sum over the whole image, where the distance between a target pixel and other pixels is employed to produce the weights. Similar pixels play a more important role in estimating the target. As for the code block, we calculate the non-local representation of a target code $y_r(p,q)$ by summarizing all the codes in $r$-th channel of the context with content-related weights. However, it is hard to determine the weights when the target code itself is unavailable for computation in the context. Given a vector $\bm \xi_r(p,q)=\{y_0(p,q),\ldots,y_{r-1}(p,q)\}$, $\bm \xi_r(p,q)$
 and $y_r(p,q)$ are produced by the same image patch. It is reasonable to evaluate the similarity between $y_r(p,q)$ and $y_r(u,v)$ with the distance between the two vectors, $\bm \xi_r(p,q)$ and $\bm \xi_r(u,v)$.  As shown in Fig.~\ref{fig:intro}(d), by finding similar code vectors, we predict the non-local representation of the target code in red with the similar codes in blue. Fig.~\ref{fig:intro} (a) shows that similar patches produce similar code vectors. As proved in many image restoration methods that employ similar data patches such as BM3D~\cite{dabov2006image}, there are many similar patches in a single image. Correspondingly, there should be many similar codes and code vectors in the code block produced by the image. Consequently, the introduced non-local operation should contribute to more accurate estimation of the entropy.

In this paper, we introduce a non-local attention block that combines the information from the local and global context for entropy modeling. Context-based convolutional networks (CCNs)~\cite{li2019efficient} are adopted to modeling the local context for their efficiency and effectiveness. In the following, we will first give a short description of the CCNs and then introduce the context-based non-local operation.

\subsection{CCNs in Modeling Local Context}
Given the code block $\bm y\in\mathbb{R}^{M\times H\times W}$, the output of the $t$-th CCN layer is a 4D tensor $\bm v^{(t)}\in \mathbb{R}^{M\times H\times W\times N_t}$ with $N_t$ feature blocks. Each feature block has the same size as the code block. The feature  $ v^{(t)}_{i,r}(p,q)$ in $r$-th channel and $i$-th feature block at spatial location $(p,q)$ is a representation of $y_{r}(p,q)$ and only convey information from the $\mathrm{CTX}(y_{i}(p,q),\bm y)$.

To speed up the efficiency and break up the serial decoding process, the CCNs divide the codes into $K=M+H+W-2$ non-overlap groups and parallel process each group, where  $\GRP_k(\bm y)=\{y_r(p,q)|r+p+q=k\}$ denotes the $k$-th group. Then, the group context is defined as $\mathrm{PTX}(y_r(p,q),\bm y)=\{y_{r'}(p', q')|r'+p'+q'<k\}$. With the group context, codes within each group share the same context and thus can be processed in parallel. Without a clear performance drop in the entropy modeling, the special group context could dramatically accelerate the decoding efficiency.

\begin{figure}[!tbp]
	\centering
	\begin{tikzpicture}
		\node[inner sep=0pt] (ctx) at (0,0)
    	{\includegraphics[width=1.0\linewidth]{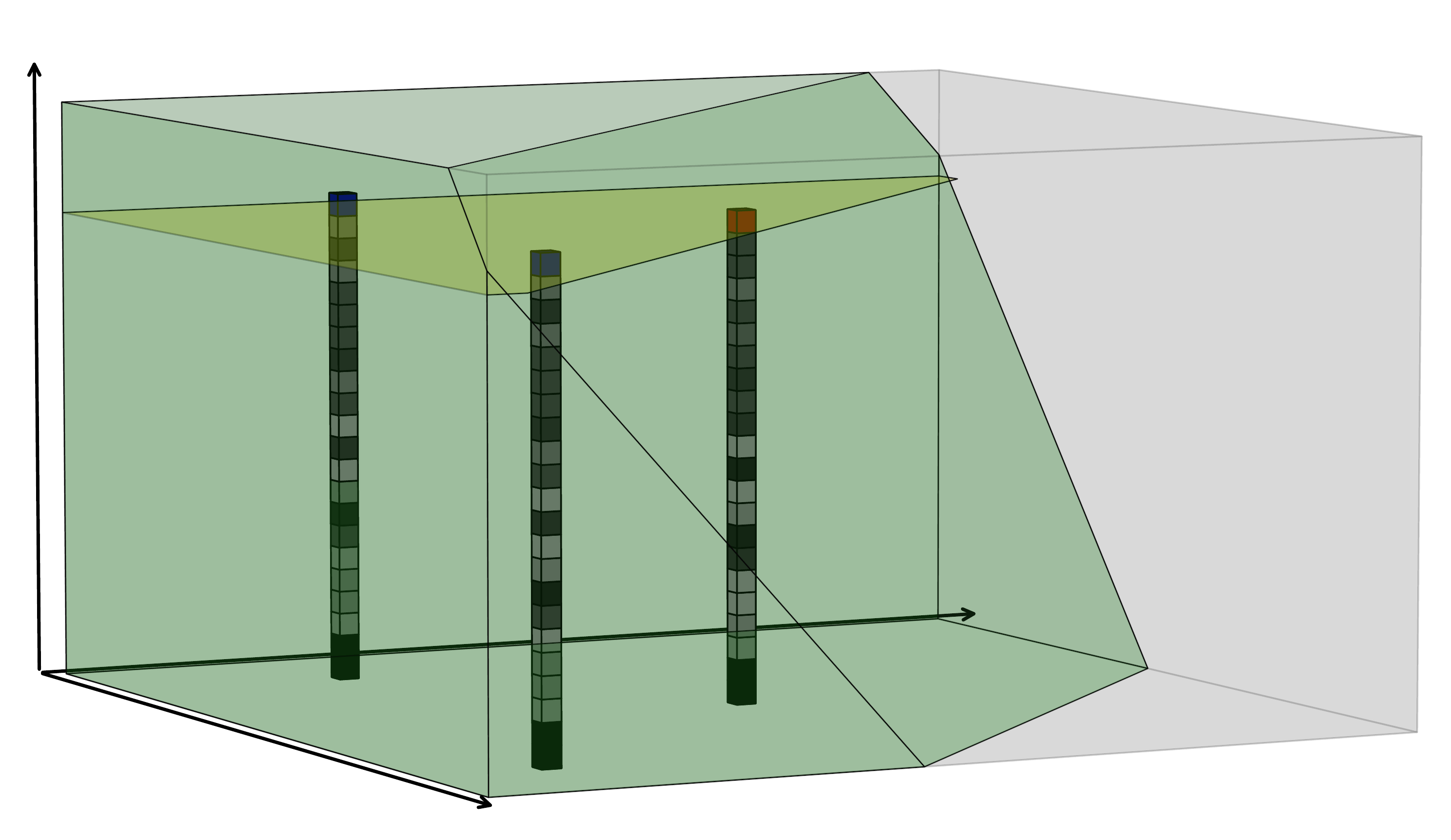}};
		\draw [line width=0.25mm,densely dotted] (0.15,1.3) -- (2.8,1.8);
		\node[text width=2cm,align=center] (cp1) at (3.5,1.8){\baselineskip=3pt \scriptsize{Target code} \par};
		\draw [line width=0.25mm,densely dotted] (0.15,-1) -- (2.5,-1);
		\draw [line width=0.25mm,densely dotted] (-1.05,-1.3) -- (2.5,-1);
		\draw [line width=0.25mm,densely dotted] (-2.35, -0.75) -- (2.5,-1);
		\node[text width=2cm,align=center] (cp2) at (3.5,-1){\baselineskip=3pt \scriptsize{Non-local similar code vectors} \par};
		\draw [line width=0.25mm,densely dotted] (-0.85,1.2) -- (2.5,0.5);
		\node[text width=2cm,align=center] (cp3) at (3.5,0.5){\baselineskip=3pt \scriptsize{Context plane for non-local operation} \par};
		
		\node[text width=0.25cm,align=center] (cp6) at (-1.2,-2.5){\baselineskip=3pt {$p$} \par};
		\node[text width=0.25cm,align=center] (cp4) at (1.8,-1.2){\baselineskip=3pt {$q$} \par};
		\node[text width=0.25cm,align=center] (cp5) at (-4.3,2.3){\baselineskip=3pt {$r$} \par};
		\draw [line width=0.25mm,densely dotted] (-1.0,1.9) -- (1.5,2.5);
		\node[text width=5cm,align=center] (cp5) at (1.5,2.7){\baselineskip=3pt \scriptsize{3D context of the code} \par};
	\end{tikzpicture}
	\caption{Illustration of the non-local similarity of the codes within the group context of CCNs for entropy modeling. The red block is the target code. And the green region is the group context for the CCNs. Content similar codes in blue are located by the proxy similarity metric $g_d$ and then employed to predict the target code. The yellow plane indicates the available codes in the non-local operation. Zoom in for better visual quality and details.}
	\label{fig:incline}
\end{figure}
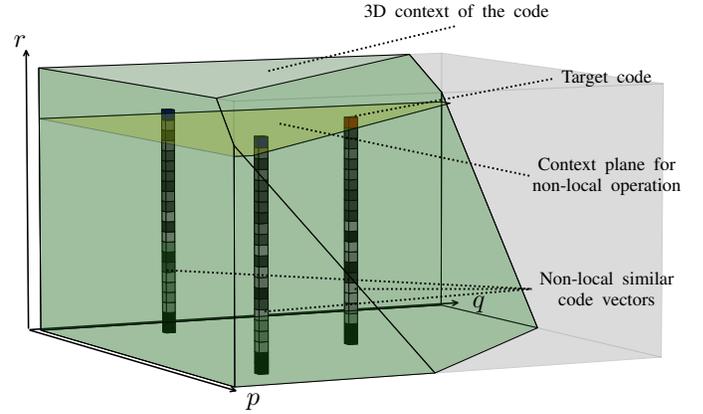

The CCNs are built on mask convolution layers which are defined as
\begin{align}
v^{(t)}_{i,r}(p,q) = \sum_{j=1}^{N_t} \sum_{s=1}^{M}\left( u^{(t)}_{j,s} \ast\left( m^{(t)}_{r,s} \odot w^{(t)}_{i,j,r,s}\right) \right)(p,q)+b_i^{(t)},
\end{align}
where $\{i,j\}$ and $\{r,s\}$ are indexes for the feature block and channel dimensions, respectively. $\bm u^{(t)}$ and $\bm v^{(t)}$ are the input and output of the $t$-th convolution layer. $\bm v^{(t-1)}$ is activated by a nonlinear element-wise function to produce $\bm u^{(t)}$. $w^{(t)}_{i,j,r,s}$ is the weight to connect feature maps $u^{(t)}_{j,s}$ and $v^{(t)}_{i,r}$, and $b_i^{(t)}$ is the bias term.
Given $w^{(t)}_{i,j,r,s}=\{w^{(t)}_{i,j,r,s}(u,v)|-k_s\leq u,v \leq k_s\}$ where $k_s$ is kernel size, the corresponding mask for the input layer is defined as
\begin{align}
	m^{(0)}_{r,s}{(u,v)} = \begin{cases}
	1, &\mbox{if } s+u+v<r\\
	0, &\mbox{otherwise}.
\end{cases}
	\label{eq:3dmask_input}
\end{align}
For the $t$-th hidden layer, the mask is modified to include the codes in the same group
\begin{align}
	m^{(t)}_{r,s}{(u,v)} = \begin{cases}
	1, &\mbox{if } s+u+v \leq r\\
	0, &\mbox{otherwise}.
\end{cases}
	\label{eq:mask_input_3d}
\end{align}

\subsection{Context-based Non-local Operation}
Different channels in the code block are generated by different convolutional filters. Only the codes in the same channel are produced by the same transform and thus should follow the same distribution. We only consider the codes in the same channel in the non-local operation.
Besides, the context should be considered.
For the target code $y_r(p,q)$, masks $\hm m^l$ are introduced to exclude the codes outside of the context.
Following the group context $\mathrm{PTX}(y_r(p,q),\bm y)$ in CCNs, we set the mask $m^l(p,q,u,v)$ to be $1$ if $y_r(u,v)\in \mathrm{PTX}(y_r(p,q),\bm y)$, \ie, $v+v<p+q$. Otherwise, $m^l(p,q,u,v)=0$.
Fig.~\ref{fig:incline} shows the codes used in the non-local operation with a yellow plane where the masks are $1$s.

Evaluating the similarity between the target and the others is necessary for non-local operation. Due to the target code is not in the context, it is unable to calculate the distance between the target and others directly.
Instead, we propose a proxy similarity function by considering the available codes in the same position but the other channels. The similarity metric $g_d(y_r(p,q),y_r(u,v))$ between the code $y_r(p,q)$ and $y_r(u,v)$ is defined as,
\begin{equation}
	g_d(y_r(p,q),y_r(u,v)) = \sum_{j=0}^{r-1} w^d_{r,j} \| y_j(p,q) - y_j(u,v) \|^2.
\label{eq:dist}
\end{equation}
Here, $w^d_{r,j}$ is the weight to balance the contribution of the proxy codes in different channels. $w^d_{r,j}$ is set to be $1/(r+1)$ in initialization and dynamically optimized with the training of the whole framework.
\begin{figure*}[!tbp]
	\centering
	\begin{minipage}{1.0\linewidth}
		\centering
		\begin{tikzpicture}
		\node[inner sep=0pt] (ctx) at (0,0)
    	{\includegraphics[width=1.0\linewidth]{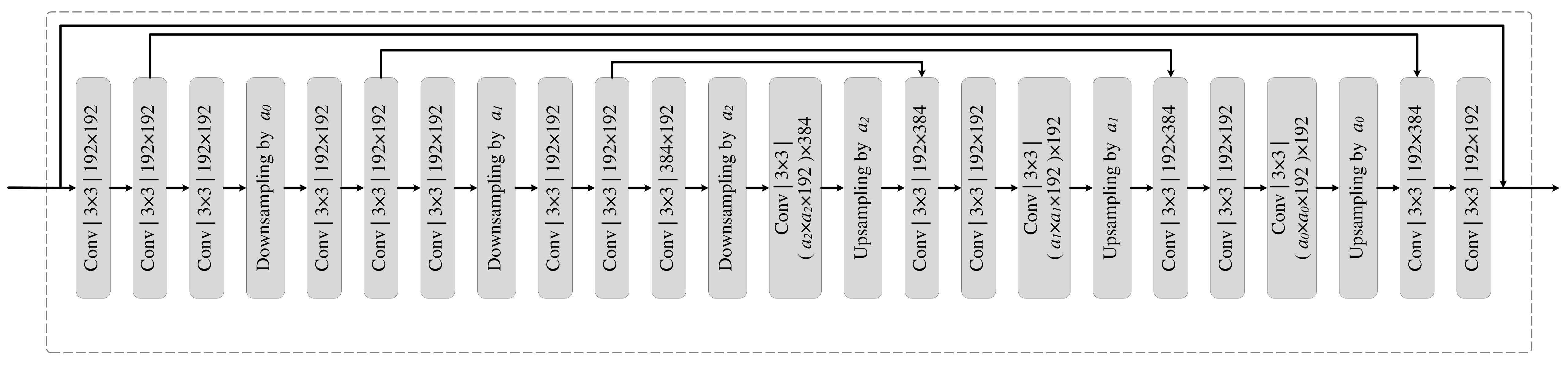}};
		\node[text width=6cm,align=center] (cp1) at (0,-1.7){\baselineskip=3pt \scriptsize{UnetBlock $|$ ($a_0$,$a_1$,$a_2$)} \par};
	\end{tikzpicture}
	\end{minipage}
	\begin{minipage}{1.0\linewidth}
		\centering{\scriptsize{(a)}}
	\end{minipage}
	\begin{minipage}{1.0\linewidth}
	\begin{tikzpicture}
		\node[inner sep=0pt] (ctx) at (0,0)
    	{\includegraphics[width=1.0\linewidth]{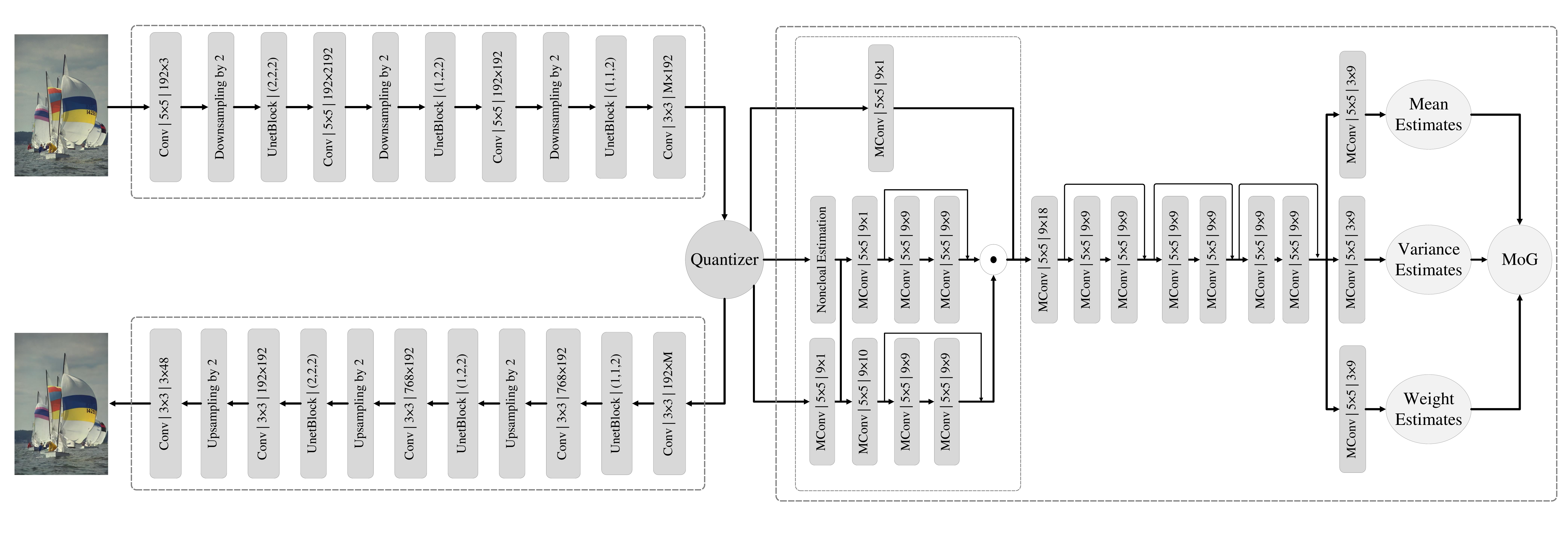}};
		\node[text width=5cm,align=center] (cp1) at (-4.2,0.6){\baselineskip=3pt \scriptsize{Analysis Transform} \par};
		\node[text width=5cm,align=center] (cp2) at (-4.2,-2.8){\baselineskip=3pt \scriptsize{Synthesis  Transform} \par};
		\node[text width=1.0cm,align=center] (cp3) at (-8.3,0.7){\baselineskip=3pt \scriptsize{Input Image} \par};
		\node[text width=1.5cm,align=center] (cp4) at (-8.3,-2.8){\baselineskip=3pt \scriptsize{Compressed Image} \par};
		\node[text width=6cm,align=center] (cp5) at (4.5,-2.9){\baselineskip=3pt \scriptsize{Context Based Non-local Entropy Model} \par};
		\node[text width=6cm,align=center] (cp5) at (1.5,-2.4){\baselineskip=3pt \scriptsize{Non-local Attention Block} \par};
	\end{tikzpicture}
	\end{minipage}
	\begin{minipage}{1.0\linewidth}
		\centering{\scriptsize{(b)}}
	\end{minipage}
	\caption{The architecture of the proposed lossy image compression method, including an analysis transform $g_a$, an adaptive trainable quantizer $g_q$, a context-based non-local entropy model $g_e$, and a synthesis transform $g_s$. Conv: regular convolution with filter support (kernel\_height $\times$ kernel\_width) and number of channels (output$\times$input).  UnetBlock: $a_0$, $a_1$ and $a_2$ are the down-sampling multipliers in the U-Net structure. MConv: mask convolution used in CCNs with filter size (kernel\_height$\times$ kernel\_width) and number of feature blocks (output$\times$input). Note that the number of channels is fixed in MConv, and is the same as the input of the entropy model, \ie, $\bm y$.}
	\label{fig:lossy_framework}
\end{figure*}
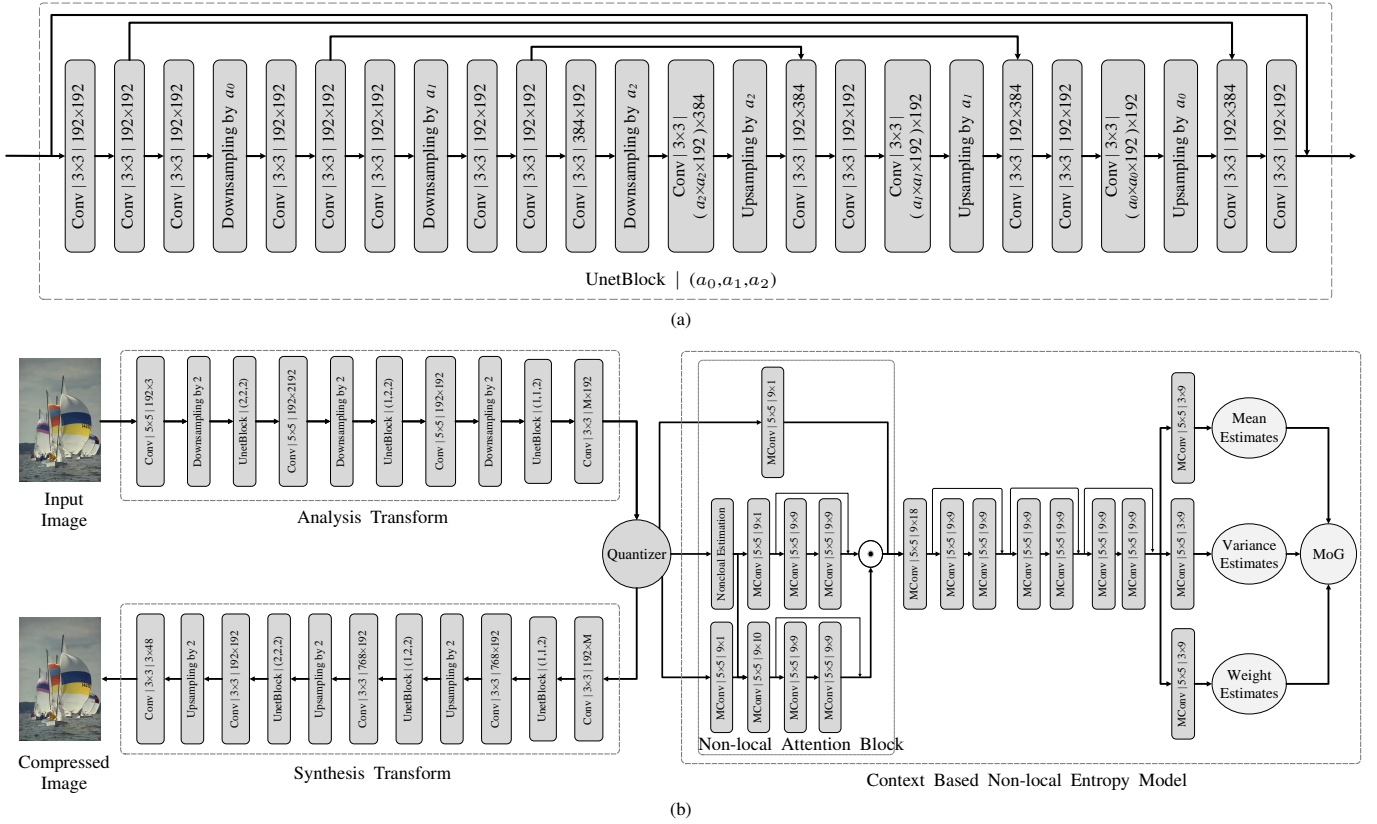
$g_d$ is a proxy based on the available codes. When considering more codes in different channels, the result will be more accurate and close to the real similarity.

The context-based non-local operation is defined as,
\begin{equation}
	g_{nlc}(y_r(p,q)) = \sum_{u,v} w^s_r(p,q,u,v) y_r(u,v),
	\label{eq:nlc}
\end{equation}
where the weight $w^s_r(p,q,u,v)$ is defined as,
\begin{equation}
	w^s_r(p,q,u,v) = \frac{m^l(p,q,u,v) e^{-g_d(y_r(p,q),y_r(u,v))}}{\sum_{u^\prime, v^\prime} {m^l(p,q,u^\prime,v^\prime) e^{-g_d(y_r(p,q),y_r(u^\prime,v^\prime))}}}.
	\label{eq:nlc_wt}
\end{equation}
An attention block is introduced to combine the non-local representation of the target code with the local representation produced by CCNs. Given the constraint of context and the proxy similarity function, the non-local representation of a target code may be predicted by the dissimilar codes and thus hinders the performance. To indicate whether the target code is estimated by content similar codes and reduce the contribution of non-local representation in those bad estimations, a confidence indicator is introduced as the weighed similarity among the proxy codes.
\begin{equation}
    g_c(y_r(p,q)) =  \sum_{u,v} w^s_r(p,q,u,v) g_d(y_r(p,q),y_r(u,v)).
    \label{eq:confidence}
\end{equation}

We combine the confidence indicators and local representation to generate the attention weight. And the Hadamard product of the attention weights and non-local representation is then concatenated with the local representation to produce output of the attention block. Fig.~\ref{fig:lossy_framework}(b) gives the structure of the block.

\section{Context-based Non-local Entropy Modeling for Lossy Image Compression}
A lossy image compression method usually involves transforms based on DNNs and is optimized for a joint rate-distortion objective function. Fig.~\ref{fig:lossy_framework} gives the framework of our lossy image compression method which consists of the  analysis transform  $g_a$, quantizer $g_q$, synthesis transform $g_s$ and the context-based non-local entropy model $g_e$. The analysis transform maps a color image $\bm x$ to the code representation $\bm z$, which is further discretized by $g_q$ to produce the code block $\bm y$. The synthesis transform $g_s$ takes $\bm y$ as input to produce a color image $\hat{\bm x}$ as the reconstruction of $\bm x$.  In this section, we will first describe the transforms, quantization function and the objective function and then introduce a post entropy model to simplify the entropy coding.
\subsection{Network Structure for Transforms}
Deep neural networks usually are non-invertible. As a result, transformations based on DNNs will lose some information in mapping the input image to the code representation. Such information loss could be ignored for models at low bit rates when the distortion between the input and the decoded image is large enough. However,  with the decrease of the distortion, the information loss brought by DNNs will start to hinder the performance. To reduce the information loss, some learned image compression methods~\cite{rippel2017real,balle2018variational,minnen2018joint} suggest adopting a wider network at high bit rates.  Consequently,  the time complexity and memory consumption also increase rapidly. A network structure taking both the computational efficiency and the width into account is needed.

U-Net~\cite{ronneberger2015u,li2018h} is a light structure originally proposed for medical image segmentation. With paired down-sampling, up-sampling operations and skip connections, U-Net is fast and can utilize representations in different scales. Besides, the skip connection could facilitate information propagation and ease the training of the network~\cite{zhang2018road}. As shown in Fig.~\ref{fig:lossy_framework}(a), we adopt the U-Net structure in building the basic block of our transforms, \ie, UnetBlock, which can keep the width and reduce the time complexity and memory consumption of the transforms at the same time. Each UnetBlock consists of 3 down-sampling and 3 up-sampling convolution layers, where the down-sampling/up-sampling multipliers are determined by $a_0$, $a_1$ and $a_2$ separately. And skip connections are introduced to combine the features at each scale. The down-sampling convolution layer is conducted by the stride convolution, while the up-sampling convolution layer is modeled as a normal convolution to increase the channels followed by the depth-to-space reshaping~\cite{toderici2016full,shi_2016_real}.

The analysis transformation $g_a$ is composed of 3 down-sampling layers with each followed by a UnetBlock. Each convolution layer is followed by a PReLU activation function. And an additional convolutional layer with sigmoid nonlinearity is adopted after the last UnetBlock to produce $\bm z \in (0,1)$ with $M$ channels. The synthesis transformation $g_s$ is a mirror of the analysis transform. And the upsampling operation is the same as used in UnetBlock. The last convolution layer makes use of three filters to produce the RGB decoded image. $\bm \phi$ and $\bm \psi$ represent the parameters to be optimized in $g_a$ and $g_s$  separately. Please refer to Fig.~\ref{fig:lossy_framework} for more details about the structure of the transforms.

\subsection{Quantization Function}

The quantization function $g_q$  maps the output of analysis transform $\bm z$ to $L$ discrete quantization centres in each channel and generates the quantized code block $\bm y$. We parametrize the quantization interval for the $r$-th channel by $\{\sigma_{r,0},\ldots,\sigma_{r,L-1}\}$. Then, the quantization centres for $r$-th channel is represented as,
\begin{equation}
	\omega_{r,i} = \sum_{j=0}^i e^{\sigma_{r,j}}, \mbox{for } i=0,\ldots,L-1.
	\label{eq:center}
\end{equation}
In quantization, the code representation $z_r(p,q)$ is assigned to the nearest quantization center in $r$-th channel.
\begin{equation}
	g_q(z_r(p,q)) = \arg \min_{\omega_{r,l}} \Vert z_{r}(p,q) - \omega_{r,l} \Vert^2_2.
	\label{eq:quant}
\end{equation}
Here, $g_q$ has zero gradients almost everywhere, which blocks the training via back-propagation. Inspired by the binarized neural networks~\cite{courbariaux2016binarized,zhou2016dorefa,rastegari2016xnor} and the DNN-based compression framework~\cite{theis2017lossy}, an identify mapping  $\hat{g}_q(z_r(p,q)) = z_r(p,q)$ is adopted as a continuous proxy for the quantization function in back-propagation.

We initialize $\bm \sigma$ as a uniform quantization in $(0,1)$, which are then optimized and updated according to the distribution of $\bm z$ by minimizing the mean squared quantization error,
\begin{align}
    \mathcal{L}_q(\bm \sigma)=\frac{1}{MHW} \sum_{r,p,q}\Vert y_{r}(p,q)-
{z}_{r}(p,q))\Vert^2_2,
    \label{eq:loss_quant}
\end{align}

\subsection{Modeling the Objective Function}

The whole framework is optimized for a joint rate-distortion objective function, where the distortion loss is directly modeled as the difference between the decoded image $\hat{\bm x}$ and the input image $\bm x$. Two separate metrics, \ie, standard mean square error (MSE) and the perceptual metric MS-SSIM~\cite{wang2003multiscale}, are adopted as the distortion loss. The MSE distortion loss $\mathcal{L}_D^{\mbox{MSE}}$ and MS-SSIM distortion loss $\mathcal{L}_D^{\mbox{MS-SSIM}}$ are defined as follows

\begin{equation}
    \mathcal{L}_D^{\mbox{MSE}}(\bm x;\bm \phi,\bm \psi) = \frac{1}{3H_IW_I}\Vert \hat{\bm x}-\bm{x}\Vert^2_2,
    \label{eq:mse}
\end{equation}
and
\begin{align}
    \mathcal{L}_D^{\mbox{MS-SSIM}}(\bm x;\bm \phi,\bm \psi) = 100- 100\;\mbox{MS-SSIM}(
    \hat{\bm x},\bm x ),
    \label{eq:ms-ssim}
\end{align}
where $\hat{\bm x}=g_s\Big(g_q\big(g_a(\bm x; \bm \phi);\bm \sigma\big);\bm \psi\Big)$. $H_I$ and $W_I$ are separately the height and width of the image $\bm x$. We denote our method optimized for $\mathcal{L}_D^{\mbox{MSE}}$ as Ours(MSE) and  $\mathcal{L}_D^{\mbox{MS-SSIM}}$ as Ours(MS-SSIM).

Each code in $\bm y$ is assumed to follow an MoG distribution depending on its context. We introduce a context-based non-local entropy model $g_e$ to produce the mixture weight, mean, and variance of these MoG distributions. As shown in Fig.~\ref{fig:lossy_framework}, the entropy model consists of a non-local attention block, several CCN residual blocks, and three final CCN layers to produce the mean, variance, and weight estimates separately. $\bm \theta$ is the parameters for the proposed entropy model.

Given $y_r(p,q)=\omega_{r,l}$, the discretized probability of a code $y_r(p,q)$ is defined as,
\begin{align}
P(y_{r}{(p,q)};\theta)=\int_{\frac{\omega_{l-1}+\omega_{l}}{2}}^{\frac{\omega_{l}+\omega_{l+1}}{2}}\sum_{i=0}^{C-1}\pi_i\mathcal{N}(\alpha; \mu_i, \sigma_i^2)d\alpha,
	\label{eq:mog}
\end{align}
where $C$ is the number of Gaussian distribution in the mixture; $\pi_i$, $\mu_i$ and $\delta^2_i$ are the mixture weight, mean, and variance of the $i$-th component, respectively. Specially, $\omega_{r,-1}=-\infty$ and $\omega_{r,L}=\infty$ for the boundary cases. Then, the entropy of the codes are adopted as the rate loss.
\begin{align}
	\mathcal{L}_R(\bm x;\bm \phi,\bm \theta) = - \sum \log P(g_q\big(g_a( x_r(p,q); \bm \phi);\bm \sigma\big),\bm \theta) .
	\label{eq:prob}
\end{align}

Finally, we propose a rate-distortion objective function for the parameters $\{\bm \phi, \bm \psi, \bm \theta\}$ over the training set $\mathcal{X}$ as
\begin{align}
	\mathcal{L}(\bm \phi, \bm \psi, \bm \theta)= \sum_{\bm x \in \mathcal{X}}\mathcal{L}_D(\bm x;\bm \phi, \bm \psi)+ \lambda \mathcal{L}_R(\bm x; \bm \phi, \bm \theta),
	\label{eq:loss}
\end{align}
where $\lambda$ is the trade-off parameter to balance the rate and distortion in the objective function.

\subsection{Post Processing for Entropy Coding}
In entropy coding, a discrete probability table that contains the probability $P(y_r(p,q)=\omega_{r,l})$ for $l=0,\ldots,L-1$ is needed. With the current entropy model built on MoG distributions, we should first produce the parameters for MoG distributions and then cut the PDF into $L$ intervals and get the discrete possibility table with an integral on each interval. Considering the amount of code to be processed is very large, it will bring further computational burden and slow down the entropy coding.
We simplify this process by training a post entropy model which takes the codes $\bm y$  as input and directly output the discrete probability table that the code belongs to each quantization center. For the network structure, we only make a small modification on the last layer. Instead of taking three separate CCN layers to produce the mean, variance, and weight estimations used in MoGs, we make use of a single CCN layer followed by a softmax nonlinearity to produce the possibility table $\bm u$ where $u_{i,r}(p,q)$ represents the possibility that $y_r(p,q) = \omega_{r,i}$. The post entropy network is optimized by minimizing the expected code length
\begin{align}
	\mathcal{L}_{post}(\bm \theta) = -\mathbb{E}_{\bm y}&\Big[\sum_{r,p,q} \sum_i {\mathbbm 1}(y_r(p,q)=\omega_{r,i})\log_2( u_{i,r}(p,q))\Big],
	\label{eq:lossless_entropy}
\end{align}
where $\mathbbm{1}(\cdot)$ is an indicator function. We first extract $\bm y$ for $\bm x \in \mathcal{X}$ with the trained analysis transform $g_a$ and then optimize the post entropy model on the extracted code blocks. Finally, we implement our own arithmetic coding with the context-based non-local entropy model to compress $\bm y$ to bitstreams, and report performance using actual bit rates.

\section{Experiments}\label{sec:exp}
In this section, we test the proposed context-based non-local entropy model and Unet-block in the lossy image compression. $10,000$ high-quality images are collected from the photo-sharing website Flickr and down-sampled to further reduce possibly compression artifacts. We crop $640,000$ color patches of size  $3\times256\times256$ as the training sets. For the post entropy model, we first extract the code blocks with the analysis transform from the full-size images and then crop code blocks with the size of $M\times60\times60$ for training. We test our models on two benchmark datasets - Kodak and Tecnick~\cite{asuni2013testimages}, and then compare them to the recent deep image compression algorithms and state-of-the-art image compression standards. The pre-trained models for testing are now available at \url{https://github.com/limuhit/Nonlocal-CCN}.

\subsection{Experimental Setup}
The quantization centers  $L$ is set to be $8$ and the number of Gaussian components in MoG  $C$ is set to be $3$. We follow the warmup strategy in~\cite{li2019efficient} and jointly optimize the transforms and entropy model by minimizing the rate-distortion objective function in Eq.~\ref{eq:loss} with an Adam solver~\cite{kingma2014adam}. Starting from a learning rate of $10^{-5}$, smaller learning rates, \ie, $10^{-6}$ and $10^{-7}$, are adopted until the loss does not decrease for 5 successive epochs.  The post entropy model is also optimized with Adam solver in the same way. We train 14 models for seven different bit rates and two distinct distortion metrics, i.e., MSE and MS-SSIM.  For testing, the compression rate is evaluated by bits per pixel (bpp), which is the total amount of bits used to compress the image divided by the whole number of pixels in the image. Two quantitative metrics, \ie, Multi-Scale Structural Similarity (MS-SSIM) and the Peak Signal-to-Noise Ratio (PSNR), are considered in evaluating the image distortion.

\subsection{Quantitative Evaluation}
Using MS-SSIM and PSNR as distortion metrics, we compare our methods with existing image compression standards such as JPEG~\cite{wallace1992jpeg}, JPEG2000~\cite{skodras2001jpeg} and BPG~\cite{bellard2016bpg} and recent DNN-based compression models in terms of rate-distortion curves.
DNN-based compression models include  Agustsson17~\cite{agustsson2017soft}, Theis17~\cite{theis2017lossy}, Toderici17~\cite{toderici2016full}, Rippel17~\cite{rippel2017real}, Mentzer18~\cite{mentzer2018conditional1}, Johnston17~\cite{johnston2017improved}, Ball\'{e}17~\cite{balle2016end}, Li18~\cite{li2017learning} and Li19~\cite{li2019efficient}.
Both JPEG (with 4:2:0 chroma subsampling) and JPEG2000 are based on the optimized implementations in MATLAB2017.
For BPG, we adopt the latest version from its official website with the default setting.
When it comes to DNN-based compression models, the implementations are generally not available.
For those models, we carefully digitalize the rate-distortion curves and report the results from their respective papers.

\begin{figure}[!tbp]
	\centering
	\begin{minipage}[t]{0.45\linewidth}
		\includegraphics[width=1.0\linewidth]{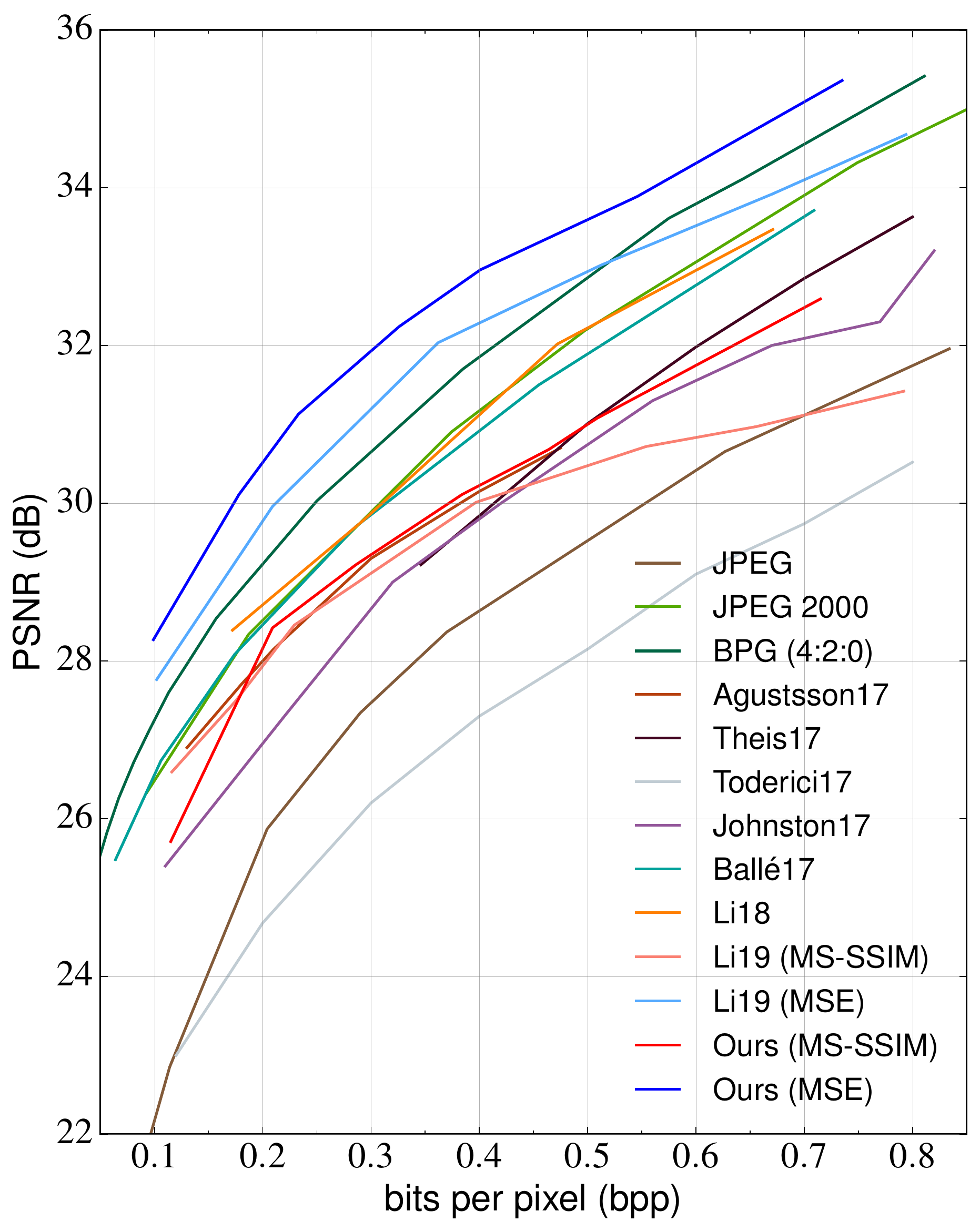}
		\centering{\scriptsize{(a)}}
	\end{minipage}
	\begin{minipage}[t]{0.46\linewidth}
		\includegraphics[width=1.0\linewidth]{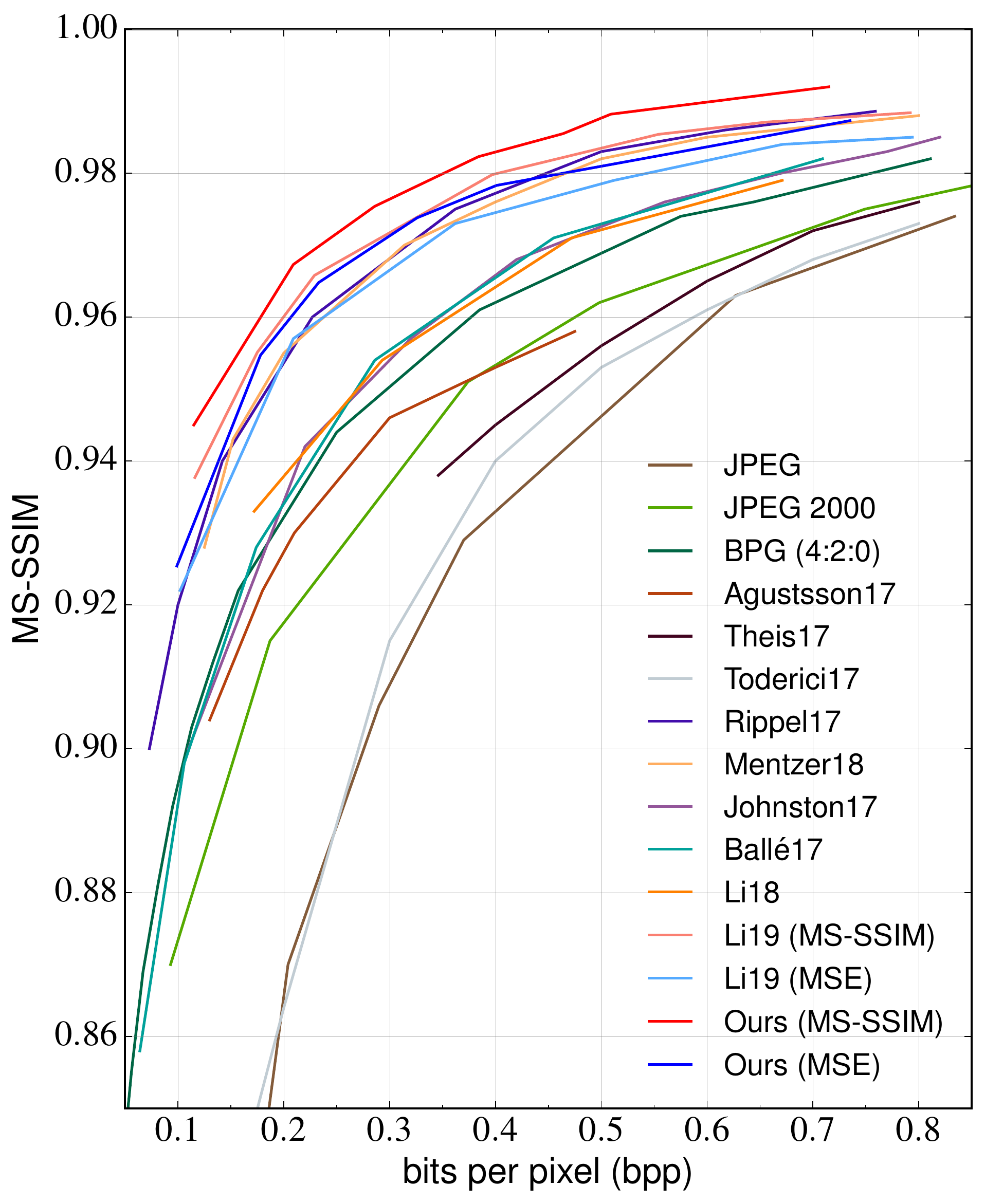}
		\centering{\scriptsize{(b)}}
	\end{minipage}
	\caption{Rate-distortion curves of different compression methods on the Kodak dataset. (a) PSNR. (b) MS-SSIM. }
	\label{fig:kodak}
\end{figure}

\begin{figure}[!tbp]
	\centering
	\begin{minipage}[t]{0.45\linewidth}
		\includegraphics[width=1.0\linewidth]{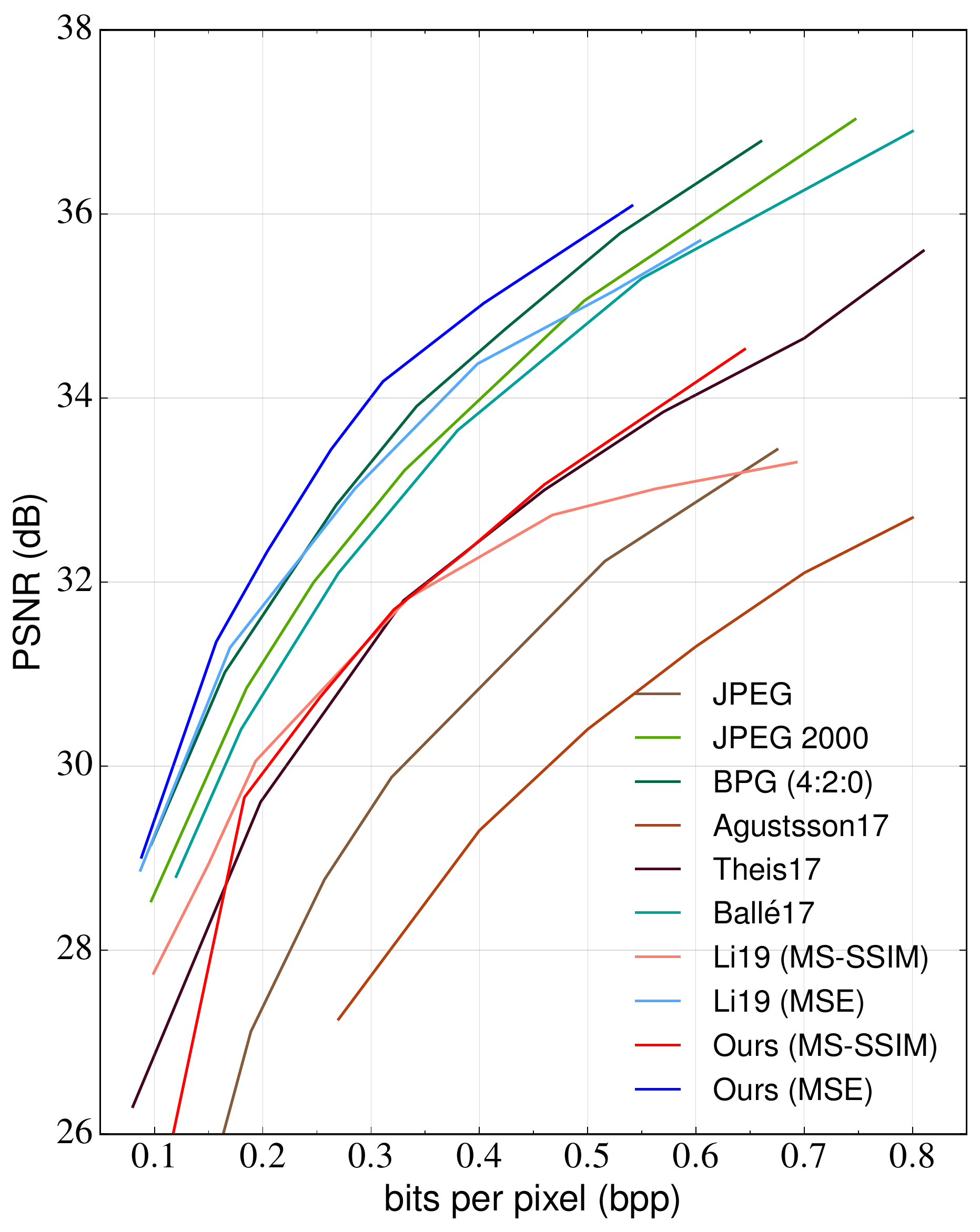}
		\centering{\scriptsize{(a)}}
	\end{minipage}
	\begin{minipage}[t]{0.46\linewidth}
		\includegraphics[width=1.0\linewidth]{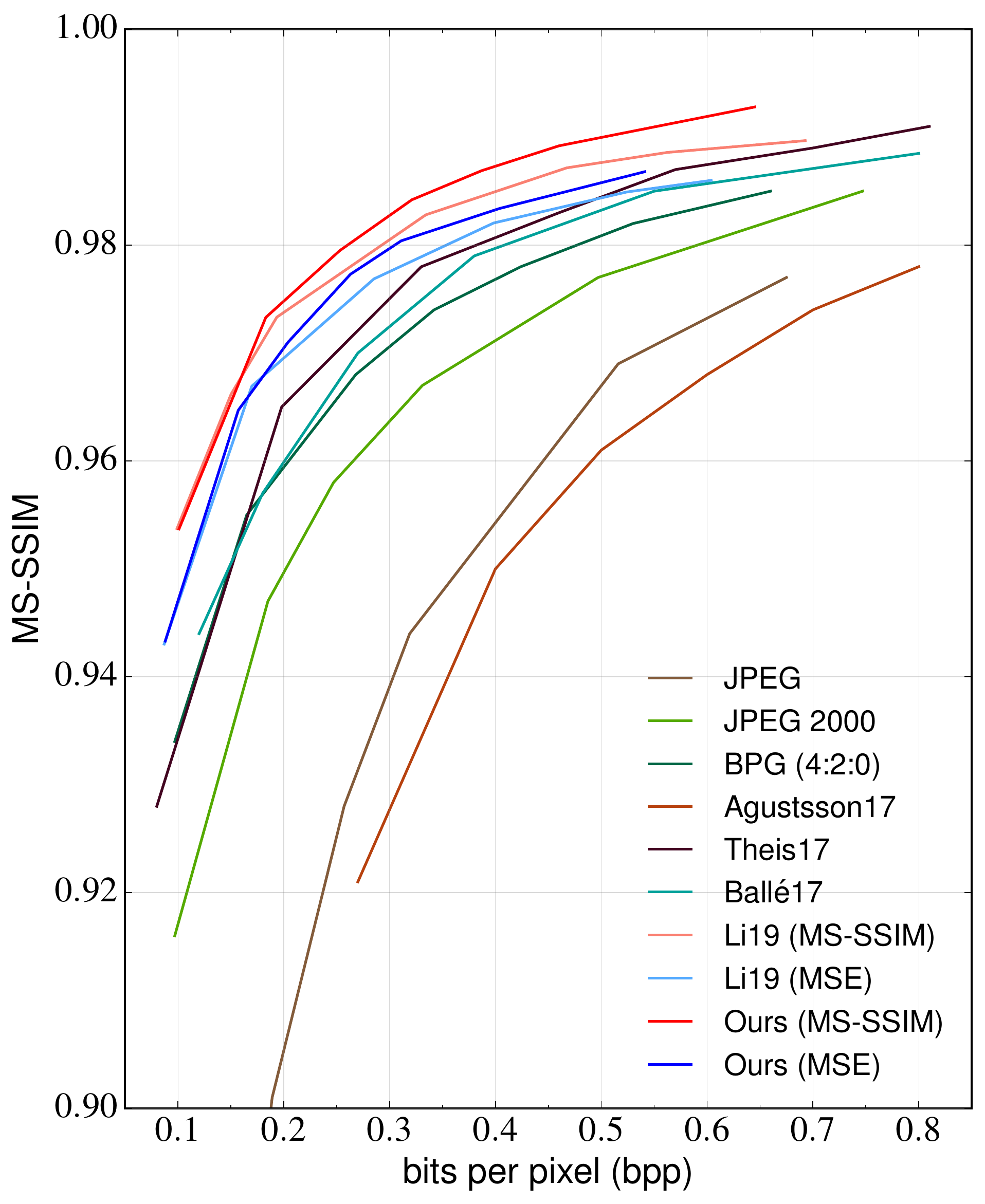}
		\centering{\scriptsize{(b)}}
	\end{minipage}
	\caption{Rate-distortion curves of different compression methods on the Tecnick dataset. (a) PSNR. (b) MS-SSIM.}
	\label{fig:tecnick}
\end{figure}

Fig.~\ref{fig:kodak} shows the rate-distortion curves on the Kodak dataset.
The results for PSNR of Rippel17~\cite{rippel2017real} and Mentzer18~\cite{mentzer2018conditional1} are missing as they do not report the PSNR curve in the paper.
We find that both Ours(MSE) and Ours(MS-SSIM) outperform all competing methods in terms of PSNR and MS-SSIM separately.
Especially, compared to Li19~\cite{li2019efficient} with local context-based entropy modeling, our models with non-local attention block are much better, which support the effectiveness of the introduced non-local operation in context modeling.
And at the high bit rate region, our models outperform Li19 by a lot, which also indicates the effectiveness of the proposed UnetBlock in reducing the information loss in transforms.
Fig.~\ref{fig:tecnick} shows the rate-distortion curves on the Tecnick dataset, where similar trends as Koadak dataset for both PSNR and MS-SSIM can be observed.
We fail to compare with Rippel17~\cite{rippel2017real}, Theis17~\cite{theis2017lossy}, Agustsson17~\cite{agustsson2017soft} and Mentzer18~\cite{mentzer2018conditional1} on Tecnick dataset due to the unavailable results in the papers.

\subsection{Visual Quality Evaluation}
\begin{figure*}
	\centering
	\begin{minipage}{1.0\linewidth}
	\includegraphics[width=1.0\linewidth]{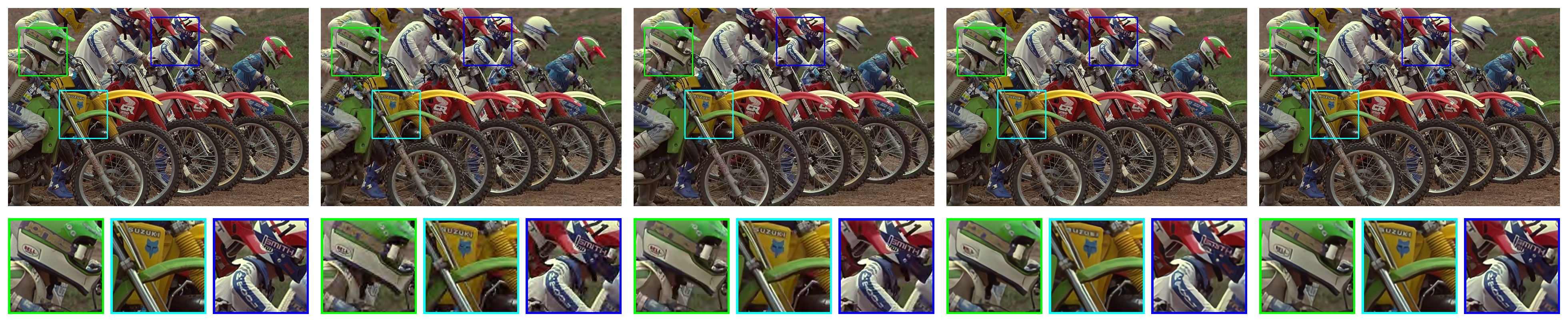}
	\end{minipage}
	\hfill\vline\hfill
	\begin{minipage}{1.0\textwidth}
	\begin{minipage}{0.2\textwidth}\center{\scriptsize{(a)}}\end{minipage}
	\begin{minipage}{0.195\textwidth}\center{\scriptsize{(b)}}\end{minipage}
	\begin{minipage}{0.195\textwidth}\center{\scriptsize{(c)}}\end{minipage}
	\begin{minipage}{0.195\textwidth}\center{\scriptsize{(d)}}\end{minipage}
	\begin{minipage}{0.195\textwidth}\center{\scriptsize{(e)}}\end{minipage}
	\end{minipage}
	\hfill\vline\hfill
	\begin{minipage}{1.0\linewidth}
	\includegraphics[width=1.0\linewidth]{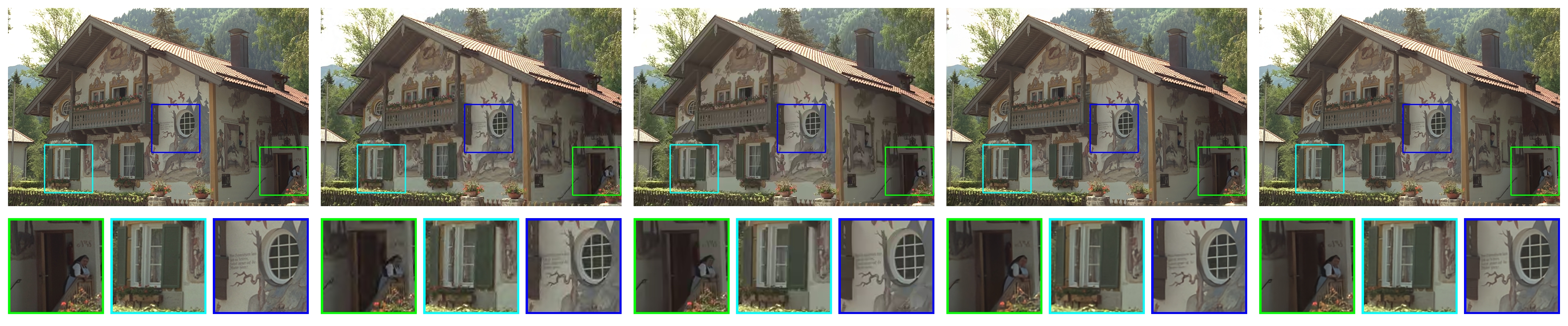}
	\end{minipage}
	\hfill\vline\hfill
	\begin{minipage}{1.0\textwidth}
	\begin{minipage}{0.2\textwidth}\center{\scriptsize{(f)}}\end{minipage}
	\begin{minipage}{0.195\textwidth}\center{\scriptsize{(g)}}\end{minipage}
	\begin{minipage}{0.195\textwidth}\center{\scriptsize{(h)}}\end{minipage}
	\begin{minipage}{0.195\textwidth}\center{\scriptsize{(i)}}\end{minipage}
	\begin{minipage}{0.195\textwidth}\center{\scriptsize{(j)}}\end{minipage}
	\end{minipage}

\caption{Compressed images by different compression methods on the Kodak dataset. The quantitative measures are in the format of ``bpp / PSNR / MS-SSIM''. (a) Uncompressed ``Motorcycle''  image. (b) JPEG2K. 0.673 / 27.75 / 0.962. (c) BPG. 0.712 / 29.65 / 0.973. (d) Li19~\cite{li2019efficient} optimized for MSE. 0.694 / 28.81 / 0.986. (e) Ours(MSE). 0.670 / 30.53 / 0.984. (f) Uncompressed ``House'' image. (g) JPEG2K. 0.879 / 31.52 / 0.972. (h) BPG. 0.877 / 32.64 / 0.979. (i) Li19 optimized for MSE. 0.871 / 29.99 / 0.988. (j) Ours(MSE). 0.865 / 33.04 / 0.989. }\label{fig:visual_kodak}
\end{figure*}

\begin{figure*}
	\centering
	\hfill
	\begin{minipage}{1.0\linewidth}
	\includegraphics[width=1.0\linewidth]{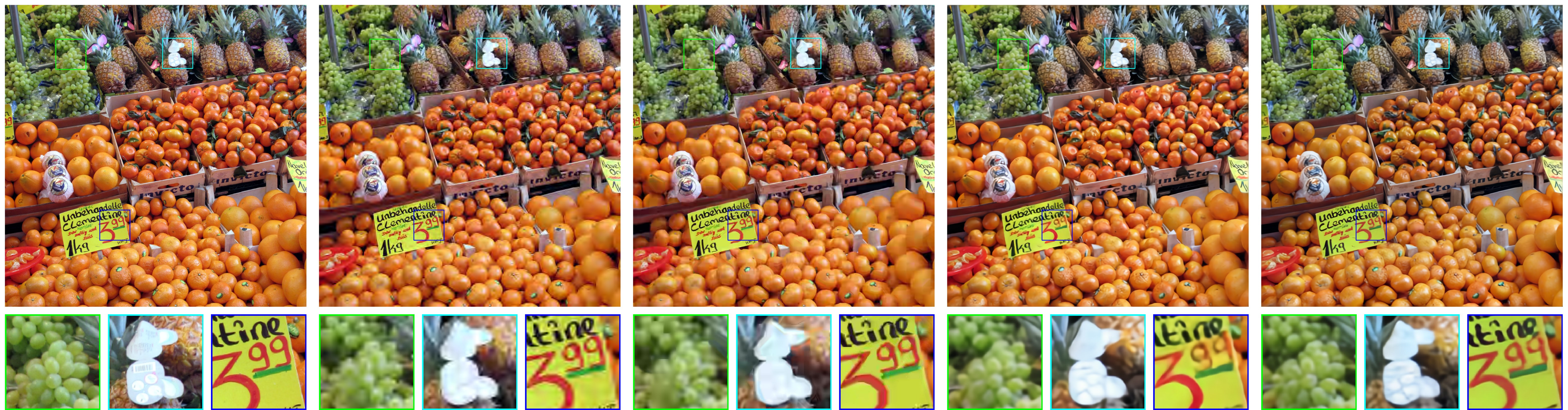}
	\end{minipage}
	\hfill\vline\hfill
	\begin{minipage}{1.0\textwidth}
	\begin{minipage}{0.2\textwidth}\center{\scriptsize{(a)}}\end{minipage}
	\begin{minipage}{0.195\textwidth}\center{\scriptsize{(b)}}\end{minipage}
	\begin{minipage}{0.195\textwidth}\center{\scriptsize{(c)}}\end{minipage}
	\begin{minipage}{0.195\textwidth}\center{\scriptsize{(d)}}\end{minipage}
	\begin{minipage}{0.195\textwidth}\center{\scriptsize{(e)}}\end{minipage}
	\end{minipage}
	\hfill\vline\hfill
	\begin{minipage}{1.0\linewidth}
	\includegraphics[width=1.0\linewidth]{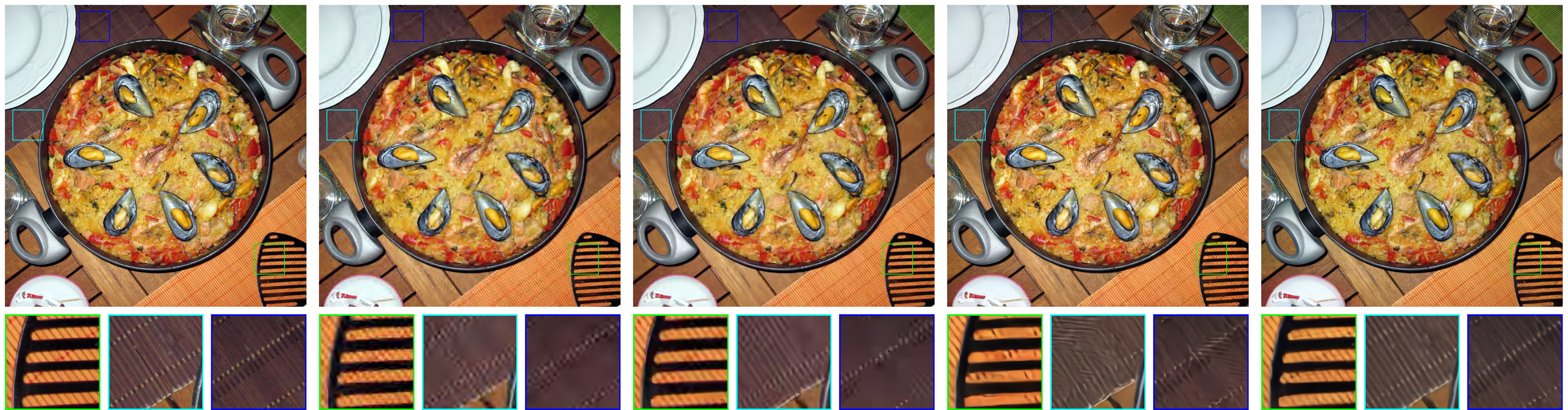}
	\end{minipage}
	\hfill\vline\hfill
	\begin{minipage}{1.0\textwidth}
	\begin{minipage}{0.2\textwidth}\center{\scriptsize{(f)}}\end{minipage}
	\begin{minipage}{0.195\textwidth}\center{\scriptsize{(g)}}\end{minipage}
	\begin{minipage}{0.195\textwidth}\center{\scriptsize{(h)}}\end{minipage}
	\begin{minipage}{0.195\textwidth}\center{\scriptsize{(i)}}\end{minipage}
	\begin{minipage}{0.195\textwidth}\center{\scriptsize{(j)}}\end{minipage}
	\end{minipage}
\caption{Compressed images by different compression methods on the Tecnick dataset. The quantitative measures are in the format of ``bpp / PSNR / MS-SSIM''. (a) Uncompressed ``Fruit'' image. (b) JPEG2K. 0.137 / 24.95 / 0.890. (c) BPG. 0.136 / 25.58 / 0.901. (d) Li19~\cite{li2019efficient} optimized for MS-SSIM. 0.139 / 25.63 / 0.938. (e) Ours(MS-SSIM). 0.134 / 24.46 / 0.939. (f) Uncompressed ``Seafood'' image.  (g) JPEG2K. 0.137 / 25.52 / 0.882. (h) BPG. 0.135 / 26.42 / 0.905. (i) Li19 optimized for MS-SSIM. 0.138 / 25.62 / 0.925. (j) Ours(MS-SSIM). 0.133 / 25.03 / 0.932.}\label{fig:visual_tec}
\end{figure*}

We further compare the decoded images by our method against Li19, JPEG2K, and BPG in visual quality and show the sample decoded images and the uncompressed images on the Kodak dataset in Fig.~\ref{fig:visual_kodak} and Tecnick dataset in Fig.~\ref{fig:visual_tec}, respectively.
For images at low bit rate, the methods optimized with MS-SSIM are visually much better due to that MS-SSIM takes structural similarity in different scale into account and are more consistent with the human visual system. When it comes to the images at a high bit rate, the methods optimized with MSE are better at keeping small scale details.
Thus, we compare Ours(MS-SSIM) with Li19(MS-SSIM) at low bit rates as shown in Fig.~\ref{fig:visual_tec} and compare Ours(MSE) with Li19(MSE) at higher bit rates in Fig~\ref{fig:visual_kodak}.
In Fig.~\ref{fig:visual_tec}, JPEG2K and BPG  exhibit artifacts (such as blocking, ringing, blurring, and aliasing) that are common to all handcrafted transform coding methods.
Li19(MS-SSIM) is effective at suppressing most of the artifacts bus still surfers from blurring in some parts of the image.
In contrast, our method optimized for MS-SSIM is more able to keep the details and has less visible distortions.
In Fig.~\ref{fig:kodak}, the whole visual quality is nearly the same. But when zooming into details, similar to BPG, our methods optimized for MSE shows to have better small scale edges and textures. Li19(MSE) blurring the small edges and losing important information, such as the text.

\subsection{Ablation Experiments}

We conduct thorough ablation experiments to analyze the impact of individual components, \ie, the UnetBlock, and context-based non-local entropy modeling, to final compression performance. For fair comparisons, we use the same set of parameters and training set for all the competing models.

\subsubsection{UnetBlock for analysis and synthesis transforms}

The width of the transforms, \ie, the number of feature maps in the output of each layer, is supposed to have a significant influence on the performance of low distortion image compression models.  A narrow transform will inevitably lose some information and thus introduce extra distortion in reconstructing the input image.  To support the effectiveness of increasing the width in low distortion cases, we introduce a baseline model by combining the entropy model from Li19~\cite{li2019efficient} and transforms based on UnetBlocks. The width for the baseline model and Li19 are $196$ and $64$.

\begin{figure}[!tbp]
	\centering
	\begin{minipage}[t]{0.45\linewidth}
		\includegraphics[width=1.0\linewidth]{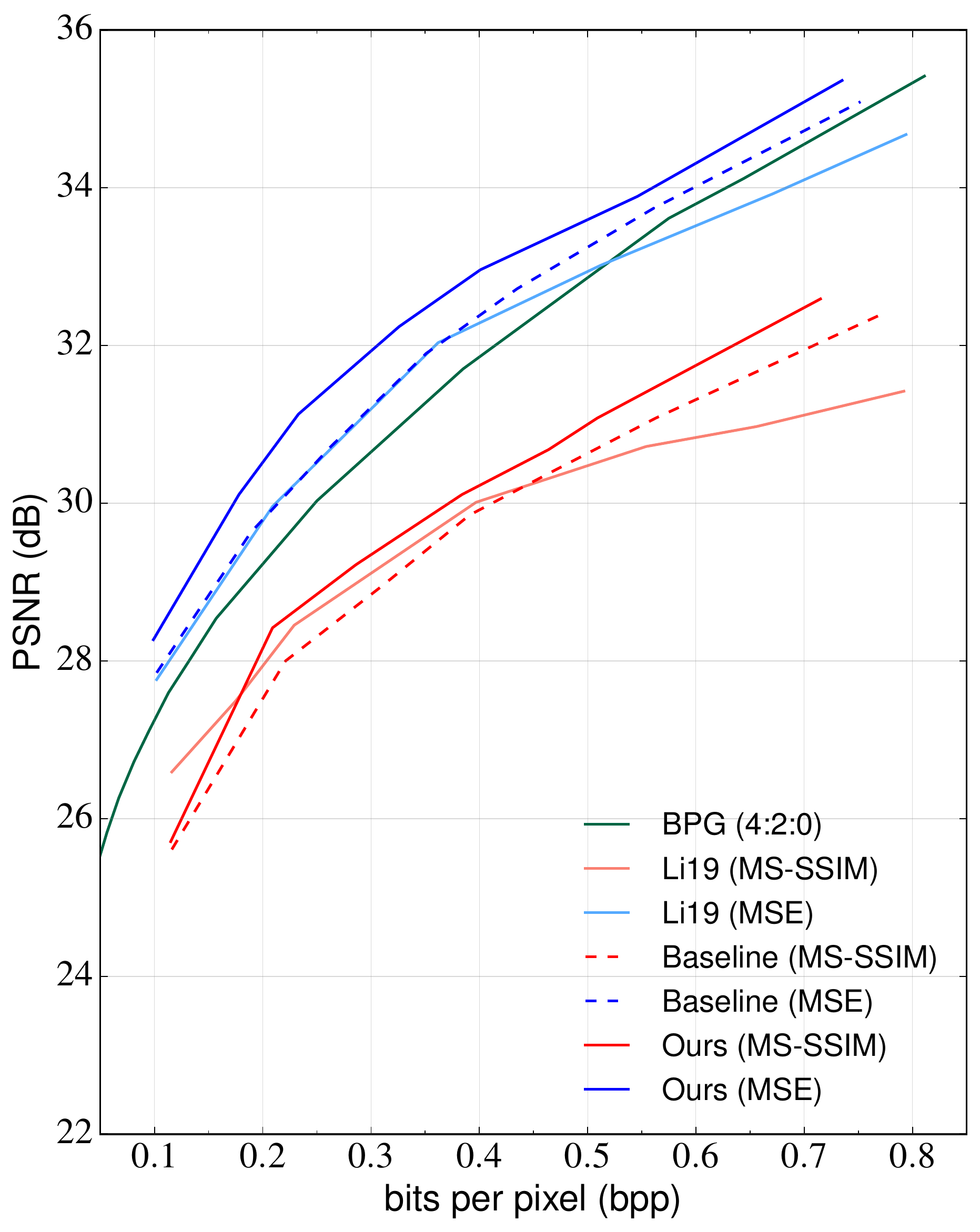}
		\centering{\scriptsize{(a)}}
	\end{minipage}
	\begin{minipage}[t]{0.46\linewidth}
		\includegraphics[width=1.0\linewidth]{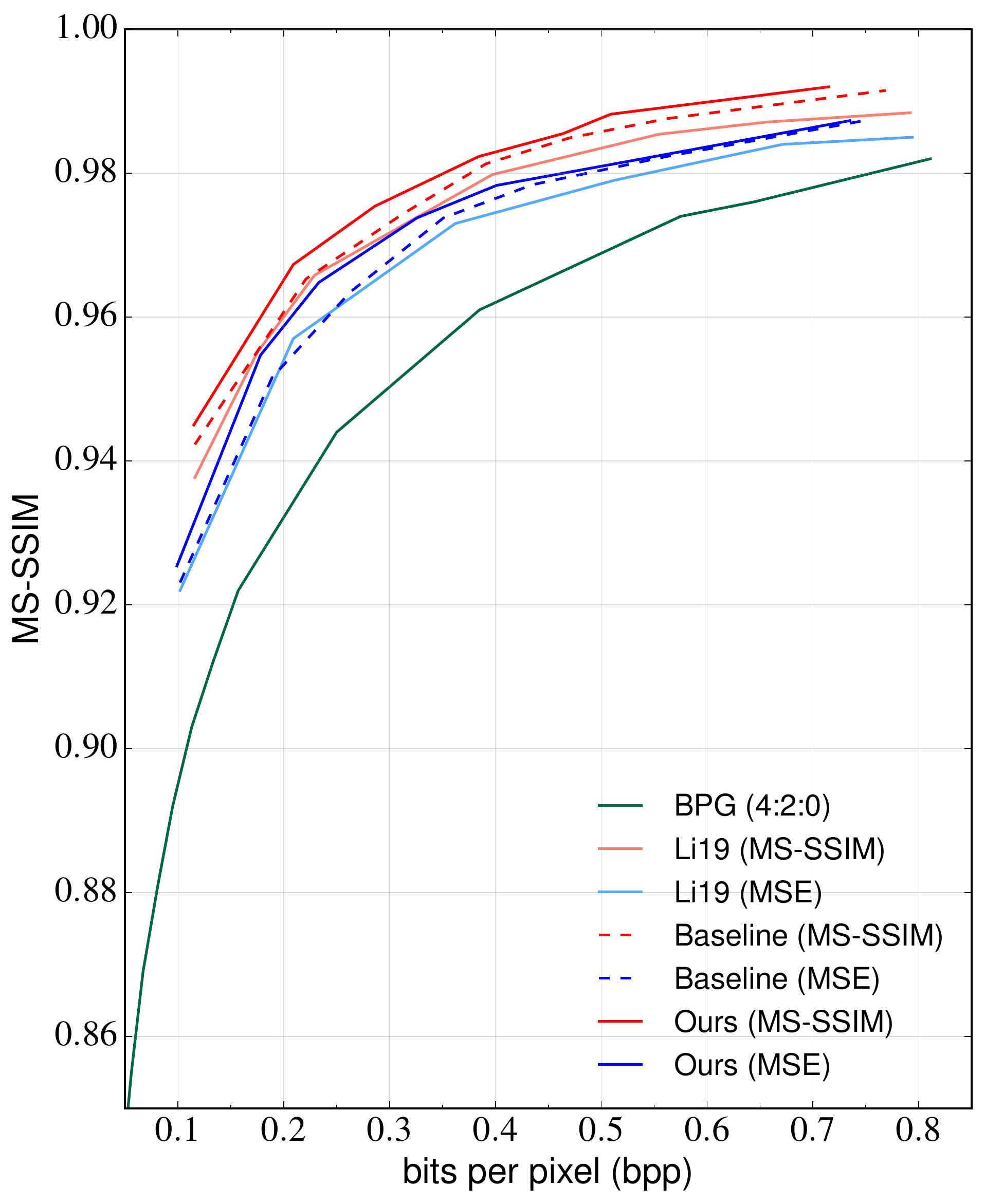}
		\centering{\scriptsize{(b)}}
	\end{minipage}
	\caption{Rate-distortion curves of different variants of the proposed methods on the Kodak dataset. (a) PSNR. (b) MS-SSIM.}
	\label{fig:ablation}
\end{figure}

\begin{table}
	\scriptsize
	\centering
	\caption{Running time in seconds, GPU memory in GBs and distortion evaluated with PSNR of three network structures. }
	\begin{tabular}{cccc}
		\toprule
	Network Structure&\specialcell{Running Time \\ (seconds)}& \specialcell{GPU Memory \\ (GB)} & \specialcell{Distortion \\ (dB)}\\
		\midrule
		DenseBlock~\cite{li2019efficient}&0.025&0.86&35.54\\
		ResidualBlock&0.136&2.65&36.34\\
		UnetBlock&0.048&1.52&36.29\\
		\bottomrule
	\end{tabular}
	
	\label{tab:performance_structure}
\end{table}

\begin{table}
	\scriptsize
	\centering
	\caption{Entropy coding for codes optimized by MSE and MS-SSIM. The results are evaluated by bits per code. { ${\bm y}^{a,i}$} ({${ \bm y}^{b,i}$}) represents the code of the $i$-th model optimized for the MSE(MS-SSIM). }
	\begin{tabular}{cccccccc}
		\toprule
		\multicolumn{8}{c}{Entropy coding for codes optimized by MSE}\\
		\midrule
		Code set & ${ \bm y}^{a,0}$ & ${ \bm y}^{a,1}$ & ${ \bm y}^{a,2}$ & ${ \bm y}^{a,3}$ & ${ \bm y}^{a,4}$ & ${ \bm y}^{a,5}$ & ${ \bm y}^{a,6}$\\
		\midrule
		CCN~\cite{li2019efficient} &1.65 & 1.56 & 1.45 &  1.45 & 1.43 & 1.46 & 1.58 \\
		Non-local&1.58 & 1.42 & 1.24 & 1.30 & 1.28 & 1.34 & 1.47\\
		\bottomrule
		\toprule
		\multicolumn{8}{c}{Entropy coding for codes optimized by MS-SSIM}\\
		\midrule
		Code set & ${ \bm y}^{b,0}$ & ${ \bm y}^{b,1}$ & ${ \bm y}^{b,2}$ & ${ \bm y}^{b,3}$ & ${ \bm y}^{b,4}$ & ${ \bm y}^{b,5}$ & ${ \bm y}^{b,6}$\\
		\midrule
		CCN~\cite{li2019efficient} &1.86 & 1.78 & 1.66 &  1.60 & 1.64 & 1.39 & 1.55 \\
		Non-local&1.84 & 1.67 & 1.53 & 1.54 & 1.48 & 1.26 & 1.43\\
		\bottomrule

	\end{tabular}
	
	\label{tab:entropy_coding}
\end{table}

\begin{figure*}
	\centering

	\begin{minipage}{0.53\linewidth}
	\begin{minipage}{1.0\textwidth}
		\includegraphics[width=1.0\linewidth]{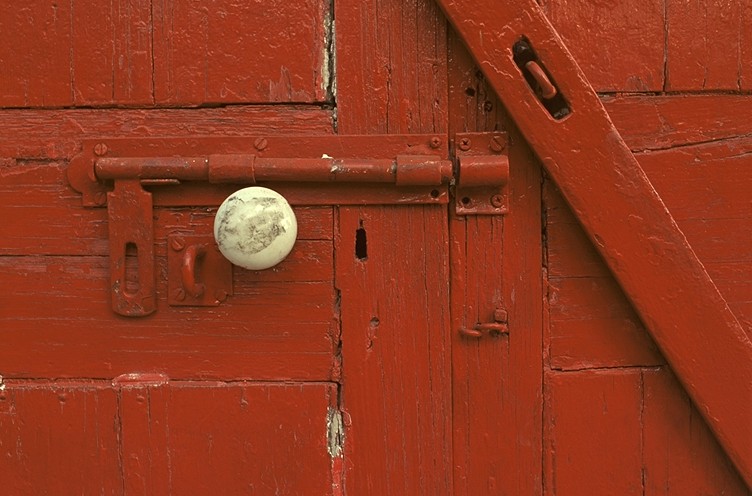}
	\end{minipage}

	\vspace{0.15cm}
	
	\begin{minipage}{1.0\textwidth}	\center{\scriptsize{(a)}}	\end{minipage}
	\end{minipage}
	\begin{minipage}{0.46\linewidth}
	\begin{minipage}{1.0\textwidth}
	\begin{minipage}{0.23\textwidth}
	\begin{minipage}{1.0\textwidth}
		\includegraphics[width=1.0\linewidth]{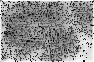}
	\end{minipage}
	
	\vspace{0.1cm}
	
	\begin{minipage}{1.0\textwidth}	\center{\scriptsize{(b)}}	\end{minipage}
	
	\vspace{0.1cm}
	
	\end{minipage}
	\begin{minipage}{0.23\textwidth}
	\begin{minipage}{1.0\textwidth}
		\includegraphics[width=1.0\linewidth]{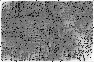}
	\end{minipage}
	
	\vspace{0.1cm}
	
	\begin{minipage}{1.0\textwidth}	\center{\scriptsize{(c)}}	\end{minipage}
	
	\vspace{0.1cm}
	
	\end{minipage}
	\begin{minipage}{0.23\textwidth}
	\begin{minipage}{1.0\textwidth}
		\includegraphics[width=1.0\linewidth]{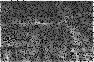}
	\end{minipage}
	
	\vspace{0.1cm}
	
	\begin{minipage}{1.0\textwidth}	\center{\scriptsize{(d)}}	\end{minipage}
	
	\vspace{0.1cm}	
	
	\end{minipage}
	\begin{minipage}{0.23\textwidth}
	\begin{minipage}{1.0\textwidth}
		\includegraphics[width=1.0\linewidth]{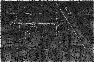}
	\end{minipage}
	
	\vspace{0.1cm}
	
	\begin{minipage}{1.0\textwidth}	\center{\scriptsize{(e)}}	\end{minipage}
	
	\vspace{0.1cm}
	
	\end{minipage}
	\end{minipage}
	\begin{minipage}{1.0\textwidth}
	\begin{minipage}{0.23\textwidth}
	\begin{minipage}{1.0\textwidth}
		\includegraphics[width=1.0\linewidth]{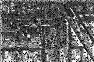}
	\end{minipage}
	
	\vspace{0.1cm}
	
	\begin{minipage}{1.0\textwidth}	\center{\scriptsize{(f)}}	\end{minipage}
	
	\vspace{0.1cm}	
	
	\end{minipage}
	\begin{minipage}{0.23\textwidth}
	\begin{minipage}{1.0\textwidth}
		\includegraphics[width=1.0\linewidth]{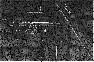}
	\end{minipage}
	
	\vspace{0.1cm}
	
	\begin{minipage}{1.0\textwidth}	\center{\scriptsize{(g)}}	\end{minipage}
	
	\vspace{0.1cm}
	
	\end{minipage}
	\begin{minipage}{0.23\textwidth}
	\begin{minipage}{1.0\textwidth}
		\includegraphics[width=1.0\linewidth]{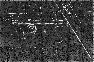}
	\end{minipage}
	
	\vspace{0.1cm}
	
	\begin{minipage}{1.0\textwidth}	\center{\scriptsize{(h)}}	\end{minipage}
	
	\vspace{0.1cm}
	
	\end{minipage}
	\begin{minipage}{0.23\textwidth}
	\begin{minipage}{1.0\textwidth}
		\includegraphics[width=1.0\linewidth]{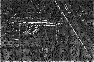}
	\end{minipage}
	
	\vspace{0.1cm}
	
	\begin{minipage}{1.0\textwidth}	\center{\scriptsize{(i)}}	\end{minipage}
	
	\vspace{0.1cm}
	
	\end{minipage}
	\end{minipage}
	\begin{minipage}{1.0\textwidth}
	\begin{minipage}{0.23\textwidth}
	\begin{minipage}{1.0\textwidth}
		\includegraphics[width=1.0\linewidth]{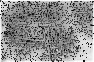}
	\end{minipage}
	
	\vspace{0.1cm}
	
	\begin{minipage}{1.0\textwidth}	\center{\scriptsize{(j)}}	\end{minipage}
	
	\vspace{0.1cm}
	
	\end{minipage}
	\begin{minipage}{0.23\textwidth}
	\begin{minipage}{1.0\textwidth}
		\includegraphics[width=1.0\linewidth]{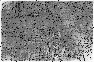}
	\end{minipage}
	
	\vspace{0.1cm}
	
	\begin{minipage}{1.0\textwidth}	\center{\scriptsize{(k)}}	\end{minipage}
	
	\vspace{0.1cm}
	
	\end{minipage}
	\begin{minipage}{0.23\textwidth}
	\begin{minipage}{1.0\textwidth}
		\includegraphics[width=1.0\linewidth]{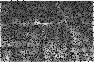}
	\end{minipage}
	
	\vspace{0.1cm}
	
	\begin{minipage}{1.0\textwidth}	\center{\scriptsize{(l)}}	\end{minipage}
	
	\vspace{0.1cm}
	
	\end{minipage}
	\begin{minipage}{0.23\textwidth}
	\begin{minipage}{1.0\textwidth}
		\includegraphics[width=1.0\linewidth]{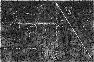}
	\end{minipage}
	
	\vspace{0.1cm}
	
	\begin{minipage}{1.0\textwidth}	\center{\scriptsize{(m)}}	\end{minipage}
	
	\vspace{0.1cm}
	
	\end{minipage}
	\end{minipage}
	\begin{minipage}{1.0\textwidth}
	\begin{minipage}{0.23\textwidth}
	\begin{minipage}{1.0\textwidth}
		\includegraphics[width=1.0\linewidth]{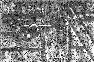}
	\end{minipage}
	
	\vspace{0.1cm}
	
	\begin{minipage}{1.0\textwidth}	\center{\scriptsize{(n)}}	\end{minipage}
	
	\vspace{0.1cm}
	
	\end{minipage}
	\begin{minipage}{0.23\textwidth}
	\begin{minipage}{1.0\textwidth}
		\includegraphics[width=1.0\linewidth]{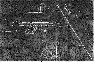}
	\end{minipage}
	
	\vspace{0.1cm}
	
	\begin{minipage}{1.0\textwidth}	\center{\scriptsize{(o)}}	\end{minipage}
	
	\vspace{0.1cm}
	
	\end{minipage}
	\begin{minipage}{0.23\textwidth}
	\begin{minipage}{1.0\textwidth}
		\includegraphics[width=1.0\linewidth]{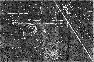}
	\end{minipage}
	
	\vspace{0.1cm}
	
	\begin{minipage}{1.0\textwidth}	\center{\scriptsize{(p)}}	\end{minipage}
	
	\vspace{0.1cm}
	
	\end{minipage}
	\begin{minipage}{0.23\textwidth}
	\begin{minipage}{1.0\textwidth}
		\includegraphics[width=1.0\linewidth]{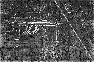}
	\end{minipage}
	
	\vspace{0.1cm}
	
	\begin{minipage}{1.0\textwidth}	\center{\scriptsize{(q)}}	\end{minipage}
	
	\vspace{0.1cm}
	
	\end{minipage}
	\end{minipage}
	\end{minipage}

\caption{Visualization of the estimated probability for the context-based non-local entropy model and the CCN-based entropy model. Each gray figure represents one code plane. The bright and dark colors indicate large and small probability of the corresponding codes. The quantitative measures are in the format of channel number of the code plane / bits per code. For simplification, "NLC" represents the context-based non-local entropy model and "LC" represents the CCN-based entropy model. (a) Uncompressed "Door" image. (b) LC. 1 / 1.05. (c) LC. 7 / 1.16. (d) LC. 16 / 1.85. (e) LC. 27 / 2.40. (f) LC. 28 / 1.67. (g) LC. 29 / 2.37. (h) LC. 30 / 2.37. (i) LC. 31 / 2.30. (j) NLC. 1 / 1.05. (k) NLC. 7 / 1.14. (l) NLC. 16 / 1.75. (m) NLC. 27 / 2.11. (n) NLC. 28 / 1.21. (o) NLC. 29 / 2.03. (p) NLC. 30 / 1.97. (q) NLC. 31 / 1.94. }\label{fig:entropy}
\end{figure*}

 We compare the baseline models separately optimized for two distortion metrics, \ie MS-SSIM and MSE, to Li19~\cite{li2019efficient} in Fig.~\ref{fig:ablation}. At a low bit rate region ($<0.4$bpp), the baseline model optimized for MSE, \ie, Baseline(MSE), and Li19(MSE) have nearly the same performance in both of the MS-SSIM and MSE. But when the bit rate grows, the Baseline(MSE) has a significant improvement, which supports the width of the transforms is important at high bit rate regions where the distortion is quite small. Similar trends can be observed for our model optimized with MS-SSIM.

With the same width and depth, Unet is usually fast in speed and needs less computational resources.  To evaluate the performance of the UnetBlock in transforms, we further compare the UnetBlock to residual block~\cite{He_2016_CVPR} on a computer with an Intel(R) Xeon(R) Processor E5-2620 v4, 64 GB of RAM and a NVIDIA TITAN Xp GPU. We adopt several residual blocks to replace a single UnetBlock to keep the depth of the competing transforms.
The DenseBlocks in Li19~\cite{li2019efficient} concatenate all the output of previous sub-blocks as the input for the next sub-block. The computational consumption increases dramatically with the growth of the width of the input to the block.
Limited by the GPU memory, we fail to compare to transforms based on DenseBlock with a width of  $192$. Thus, we only adopt the transforms with a width of $64$ for DenseBlock in comparison.
Without the limitation of the entropy model ($\lambda = \infty$), the three competing models are optimized for the MSE distortion loss.
Table~\ref{tab:performance_structure} gives the results of the three structures on running time in seconds, GPU memory in GB and distortion performance evaluated by PSNR. The transforms based on the ResidualBlock can not process the whole image from the Kodak dataset due to the GPU memory. We only report the performance on $256\times256$ patches sampled from the Kodak dataset.
The UnetBlock performs on par with ResidualBlock on distortion performance and overwhelm the narrow network, \ie, DenseBlock.
When it comes to efficiency, the transforms based on UnetBlock is much faster and needs less GPU memory than the transforms based on ResidualBlock.
Thus, the UnetBlock is a good trade-off between the speed and performance.

\subsubsection{Context-based non-local entropy modeling}
The context-base non-local entropy modeling exploits both of the global similarity and the local representations of the context for entropy modeling. We compare it to the entropy model only makes use of local representations~\cite{li2019efficient}. The same baseline models are used for comparison. Compared to Ours(MSE) and Ours(MS-SSIM), the baselines adopt the CCN-based entropy models which focus on local representations. As shown in Fig.~\ref{fig:ablation}, both Ours(MSE) and Ours(MS-SSIM) outperform the counterpart baseline models by a large margin, which strongly supports the effectiveness of the proposed non-local attention block in context modeling.

To remove the effectiveness of different initialization and training process,  we test the performance of the context-based non-local entropy models and the CCN-based entropy models on the extracted codes. ${\bm y}^{a,0}, \ldots, {\bm y}^{a,6}$ denote seven code sets produced by Ours(MSE) at seven different bit rates.   ${\bm y}^{b,0}, \ldots, {\bm y}^{b,6}$ are produced by Ours(MS-SSIM). The performance of the two entropy models is evaluated by bits per code, which is the rate between the total bits used to coding the code block and the number of codes in the code block. Table~\ref{tab:entropy_coding} gives the performance of the two entropy models, the context-based non-local entropy model is significantly better than the CCN-based entropy model on the 14 code sets, which supports the contribution of the introduced non-local attention block in entropy modeling.

We further visualize the estimated probability for each code plane by mapping $P(y_r(p,q))$ to an integer in the range of $[0,255]$ and shows the code plane as a gray image in Fig.~\ref{fig:entropy}.
Fig.~\ref{fig:entropy} (a) gives the uncompressed image; (b)-(i) are the visualized probability of the CCN-based entropy model on 8 code planes; (j)-(q) are the corresponding results of the context-based non-local entropy model.
In bottom code planes, as shown in Fig.~\ref{fig:entropy} (b), (c), (j) and (k), CCN-based entropy model has similar performance as the proposed non-local entropy model.
This can be illustrated by the proxy similarity function.
The similarity of codes inner one plane is estimated by the proxy codes in the bottom code planes. With a few numbers of bottom code planes, the similarity is not accurate and thus lead to poor non-local estimation.
The attention block tends to focus on the local representations instead of the non-local estimation. Thus, the competing entropy models share a similar performance.
With the increase of planes in the bottom, the proxy similarity begins more accurate and the non-local estimations star to contribute to the performance.
The context-based non-local entropy modeling shows to have overwhelming performance in the top code planes as described in Fig.~\ref{fig:entropy} (f), (g), (h), (i), (n), (o), (p) and (q).

\section{Conclusion}
In this paper, we solved the missing reference problem in the context modeling with a proxy similarity metric and introduced the non-local operation for contex-based entropy modeling.
With a non-local attention block to combine the local and global representations, our non-local entropy models can provide more precise estimation for the entropy of the codes.
To reduce the information loss and boost the performance for low distortion compression, we further designed an effective and efficient network structure, \ie, UnetBlock, and adopted it in building the transforms of the proposed lossy image compression framework.
We tested the context-based non-local entropy model and the UnetBlock in lossy image compression.
Both of our models optimized for MSE and MS-SSIM got significant improvements over state-of-art image compression standards and recent DNN-based models.


\section*{Acknowledgment}
The authors would like to thank the NVIDIA Corporation for donating a TITAN Xp GPU used in this research.


\bibliographystyle{IEEEtran}
\bibliography{IEEEabrv,./egbib}
\end{document}